\RequirePackage{lineno}
\documentclass[superscriptaddress,aps,preprint,amsmath,amssymb,floatfix]{revtex4-1}

\usepackage{latexsym}
\usepackage{lineno}
\usepackage{hyperref}
\usepackage{url}
\usepackage[caption=false]{subfig}
\usepackage{graphicx}
\usepackage{verbatim}
\usepackage{multirow}
\usepackage{amsmath, bm}
\newcommand{\beq}{\begin{eqnarray}}
\newcommand{\eeq}{\end{eqnarray}}
\usepackage{mathrsfs}
\usepackage{float}
\usepackage{dsfont}
\usepackage{bbm}
\usepackage[usenames, dvipsnames]{color}
\usepackage{mathtools}
\usepackage{physics}	
\usepackage{graphicx}  
\usepackage{epstopdf}
\usepackage{hyperref}   
\usepackage{bbold}
\usepackage{wasysym}
\usepackage{amssymb}
\usepackage{feynmp}
\usepackage[symbol]{footmisc}
\usepackage[latin1]{inputenc}
\usepackage{tikz}
\usetikzlibrary{decorations.pathmorphing}
\usetikzlibrary{shapes.misc}
\tikzset{cross/.style={cross out, draw=black, minimum size=8*(#1-\pgflinewidth), inner sep=0pt, outer sep=0pt},
cross/.default={1pt}}
\usetikzlibrary{patterns,math}
\newcommand{\RN}[1]{%
  \textup{\uppercase\expandafter{\romannumeral#1}}%
}
\usepackage{bm}
\usepackage{gensymb}

\newcommand{\be}[0]{\begin{equation}}
\newcommand{\ee}[0]{\end{equation}}
\newcommand{\ba}[0]{\begin{eqnarray}}
\newcommand{\ea}[0]{\end{eqnarray}}
\newcommand{\mx}[0]{\begin{pmatrix}}
\newcommand{\ex}[0]{\end{pmatrix}}

\newcommand{\ncmd}{\newcommand}
\ncmd{\nn}{\nonumber}
\ncmd{\mbf}[1]{\bs{#1}}
\ncmd{\gam}{\gamma}
\ncmd{\sig}{\sigma}
\ncmd{\pha}{\alpha}
\ncmd{\lam}{\lambda}
\ncmd{\dl}{\delta}
\ncmd{\kap}{\kappa}
\ncmd{\Lam}{\Lambda}
\ncmd{\Gam}{\Gamma}
\ncmd{\Dl}{\Delta}
\ncmd{\Ups}{\Upsilon}
\ncmd{\Om}{\Omega}
\ncmd{\eps}{\epsilon}
\ncmd{\veps}{\varepsilon}
\ncmd{\vphi}{\varphi}
\ncmd{\vtheta}{\vartheta}
\ncmd{\tw}{\text{w}}
\ncmd{\pll}{\parallel}
\ncmd{\mc}{\mathcal}
\ncmd{\mf}{\mathfrak}
\ncmd{\bs}{\boldsymbol}
\usepackage{pifont}
\usepackage[T1]{fontenc}
\usepackage{bclogo}
\ncmd{\note}[1]{{\color{red}{\ding{168} #1}}}
\ncmd{\eq}[1]{Eq. \eqref{#1}}
\ncmd{\fig}[1]{Fig. \ref{#1}}
\ncmd{\suppl}{\note{`Supplementary Information'}}
\ncmd{\pg}[1]{\textcolor{red}{#1}}

\begin{document}

\hyphenation{va-ni-sh-ing}

\begin{center}

\thispagestyle{empty}

{\large\bf Topological Diagnosis
of Strongly Correlated Electron Systems}
\\[0.3cm]

Chandan Setty$^{1,\ast,\dagger}$, Fang\ Xie$^{1,\dagger}$, Shouvik Sur$^{1,\dagger}$, Lei\ Chen$^{1}$,  
 Silke\  Paschen$^{2,1}$, \\
 Maia\ G.\ Vergniory$^{3,4}$, Jennifer Cano$^{5,6}$, and Qimiao Si$^{1,\ast}$
\\[0.2cm]

$^1$Department of Physics and Astronomy, Rice Center for Quantum Materials, Rice University, Houston, Texas 77005, USA
\\[-0.cm]

$^2$Institute of Solid State Physics, Vienna University of Technology, Wiedner Hauptstr. 8-10, 1040, Vienna, Austria
\\[-0.cm]

$^3$Donostia International  Physics  Center,  P. Manuel  de Lardizabal 4,  20018 Donostia-San Sebastian,  Spain
\\[-0.cm]

$^4$Max Planck Institute for Chemical Physics of Solids, Noethnitzer Str. 40, 01187 Dresden, Germany
\\[-0.cm]

$^5$Department of Physics and Astronomy, Stony Brook University, Stony Brook, NY 11794, USA
\\[-0.cm]

$^6$Center for Computational Quantum Physics, Flatiron Institute, New York, NY 10010, USA

\end{center}

\vspace{0.08cm}
{\bf 
The intersection of electronic topology and strong correlations offers a rich platform to discover exotic
quantum phases of matter and 
unusual 
materials.
An overarching challenge 
that impedes the discovery is how to
diagnose 
topology in strongly correlated settings,
as exemplified by Mott insulators.
Here, we  develop a general framework to address this outstanding question 
and illustrate its power in the case of Mott insulators.
The concept of
Green's function Berry curvature---which is frequency dependent---is introduced.
We apply this notion in
a system that contains symmetry-protected nodes in its noninteracting bandstructure;
strong correlations drive the system into a Mott insulating state,
creating contours in frequency-momentum space where the Green's function vanishes. 
The Green's function Berry flux 
of such zeros
is found to be quantized,
and is as such direct probe of
the system's topology.
Our framework allows for a comprehensive search of strongly correlated topological
materials with Green's function topology.
}

\vspace{0.6cm}

\noindent $^\ast$ E-mails: csetty@rice.edu; qmsi@rice.edu

\clearpage 

 Strong correlations can drive new topological phases of matter. 
 This happens in insulators 
 as exemplified by the
fractional quantum Hall effect \cite{Stormer99} and
as evinced
in emerging fractional Chern insulators \cite{Xie2021,Xu2023,Zeng-fci2023},
and is being realized in
metallic systems such as 
the Weyl-Kondo semimetals in heavy fermion models \cite{Lai2018,Chen-Si2022} and 
materials \cite{Dzsaber2017,Dzs-giant21.1} with nonsymmorphic 
 symmetries.
An outstanding question is
whether and how Mott insulators can be topological. 
The central role that electron correlations play in the 
formation of Mott insulators suggests that novel types of topological phases may develop in these systems. The Mott gaps
are necessarily large, 
suggesting that any topological phase will be 
especially robust. However, there is presently no methodology to discover such phases in real materials.
To make progress,
approaches to diagnose electronic topology in such settings are urgently needed.

\begin{figure}[t!]
    \centering
    \includegraphics[width=1\linewidth]{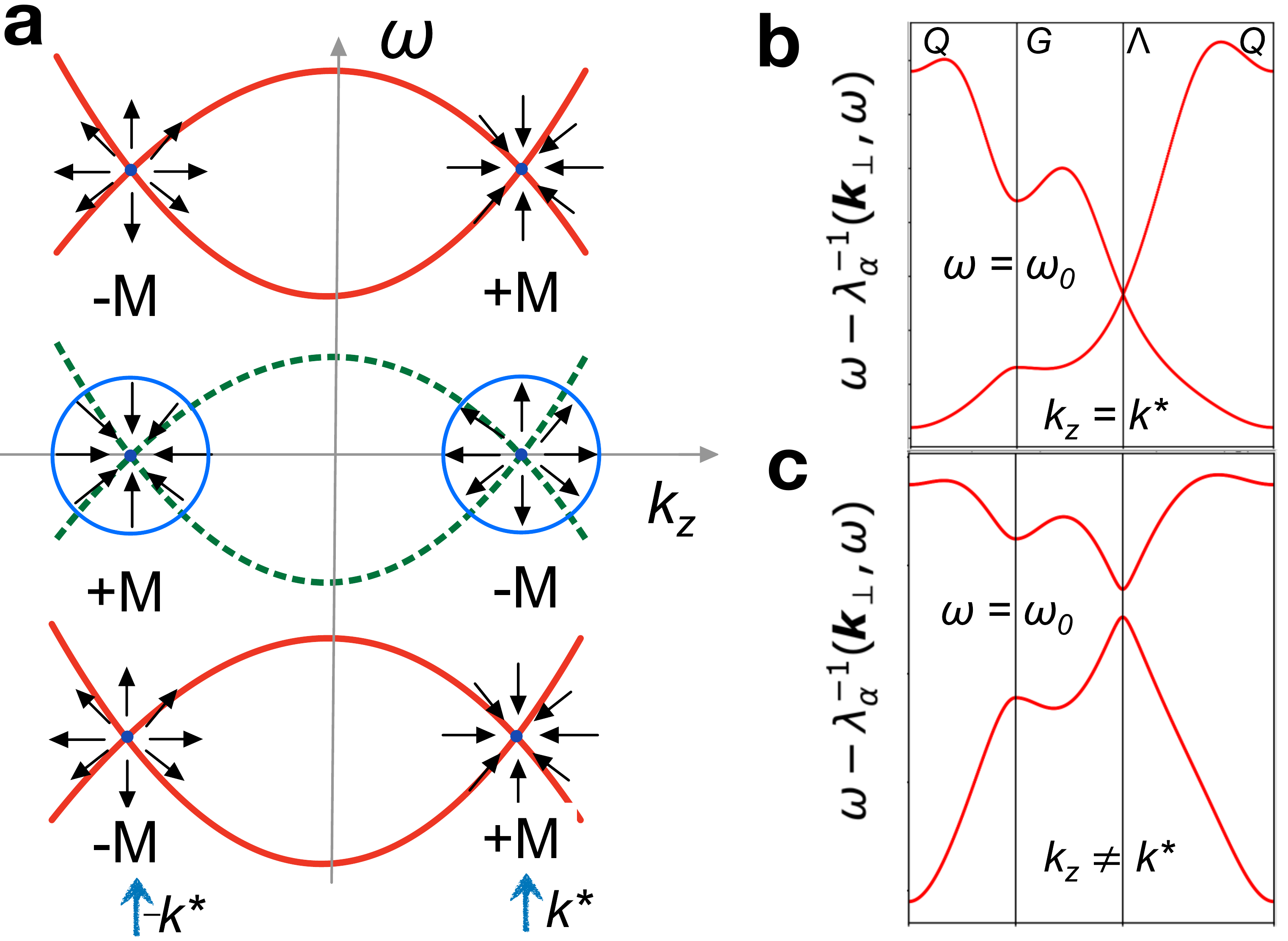}
    \caption{\textbf{Berry flux quantization and Green's function eigenvalues. }Schematic plots of 
    \textbf{(a)} 
     retarded Green's function pole (red solid lines) and zero 
     (green
     dashed lines) crossings as a function of $k_z$ for a fixed $(k_x, k_y)$. The crossings behave as sources and sinks of Berry flux (black arrows computed using $G_+(\bs k,\omega)$ eigenvectors in Eq.~\ref{Eq:BerryCurvature}). 
    The blue circles represent Gaussian spheres from which one can compute the frequency-dependent Chern number of poles and zeros resulting in a non-trivial quantization condition. The enclosed quantized topological charge is denoted by $\pm \text M$. The blue arrows denote $k_z = \pm k^*$ at which there is spectral crossing. \textbf{(b)}  $G_+(\bs k, \omega)^{-1}$ `bands' (or eigenvalues of $G_+(\bs k, \omega)^{-1}$) as a function of $k_{\perp} = (k_x, k_y)$ shifted by $\omega$ along high symmetry lines for a fixed $(k_z = k^*, \omega = \omega_0)$. $\Lambda, G, Q$ are high symmetry lines along the $k_z$ direction which become points for a fixed $k_z$. See SI Sec.~\ref{app:lattice} Fig.\ref{fig:lattice}(b) for the Brillouin zone. There are two pairs of doubly degenerate `bands'. The bottom pair is  used to compute the Berry curvature. \textbf{(c)} Same as \textbf{(b)} but for $k_z \neq k^*$.
    }
    \label{ZeroMonopole}
\end{figure}

Here, we introduce a framework that defines and computes the Berry curvature in interacting settings using 
Green's function eigenvectors.
This Green's function Berry curvature
is frequency dependent.
We demonstrate
that the Berry flux generated when two zeros in the Green's function cross in momentum space
encloses a quantized monopole Berry charge (see schematic Fig.~\ref{ZeroMonopole}(a)). 
 Our appraoch, therefore, provides a systematic means
 to diagnose 
 electronic
 topology in strongly correlated
systems. 
Consequently, the proposed framework realizes the first  
route to search for correlated materials with topological zeros.

\medskip

\noindent
{\bf Green's function and its zeros}

The Green's function is the propagator of 
an electron in time and space
\cite{abrikosov2012}.
For noninteracting electrons in a crystal, its Fourier transform into frequency and wavevector displays a pole when viewed for a fixed wavevector as a function of frequency. This pole captures Bloch states, the eigenstates of 
an electron moving in the periodic potential of the crystal. 
Here, the crystalline symmetry constrains the energy dispersion. For example, it allows the identification and classification of band degeneracies enforced or protected by symmetry, which lead to nodes in the energy dispersion \cite{Armitage2017,Nagaosa2020, Bradlyn2017,Cano2018,Po2017,Watanabe2017,cano2021band} and the emergence of topological semimetals.

One way to treat the interactions is through the approach of 
a topological Hamiltonian,
 which describes an interacting system in terms of a noninteracting one with a single-particle Hamiltonian renormalized by the $\omega=0$ component 
of the self-energy~\cite{Wang-Zhang_PRX2012, Wang-Yan2013, Iraola21, Lessnich2021,Soldini2022}.
For Green's function zeros, though, the self energy diverges at $\omega=0$ (or, more generally, at the location of zeros), and the topological Hamiltonian approach cannot be applied.

The approach advanced here rests on the recent realization that 
Green's function eigenvectors 
form a representation of the lattice space group similar to 
what the Bloch functions
do in noninteracting systems~\cite{Hu-Si2021}.
As such, the Green's function eigenvectors provide a means to implement constraints of space group symmetries on single-particle excitations regardless of whether 
quasiparticles exist~\cite{Hu-Si2021,Setty2023}.
This formulation allowed for the identification of Kondo-driven Weyl semimetals 
without well-defined
quasiparticles, with Weyl nodes that are specified in terms of spectral crossing, i.e. a degeneracy in the Green's function eigenvectors~\cite{Hu-Si2021}.
Indeed, the lattice symmetry 
operates on the Green's function zeros \cite{Setty2023}.

There is in addition an increasing recognition that
zeros 
participate in 
the electronic topology~\cite{AGD, Dzyaloshinskii2003, sym_green, Gurarie2011-2, Yunoki2017, Setty2023, SettySi2023b, wagner2023mott, Fabrizio2022,Fabrizio2023}.
As such, identifying materials with Green's function zeros is a key goal of realizing 
topological 
Mott insulating materials.
We note that 
electronic topology has been
analyzed in a Hartree-Fock treatment of interactions for Mott insulators, in which zeros do not appear~\cite{Raghu-Zhang2008}. Furthermore, Mott insulators have been studied in the context of Weyl semimetals~\cite{Nagaosa2016} where zeros 
appear at a single point
(i.e., they do not disperse). 
More recently, Mott phases with dispersive zeros were suggested to occur in moir\'e materials~\cite{Phillips2023-NatCommun}.

\medskip

\noindent
{\bf Berry phase}

An essential concept 
for realizing electronic topology 
is the 
Berry phase~\cite{Pancharatnam1956, Berry1984}, a geometric phase that is acquired along a closed path within a parameter space of an adiabatically evolving system. For noninteracting electrons, any many-body wavefunction can be expressed as a symmetrized product of the Bloch states. 
These Bloch functions acquire a Berry phase \cite{ Zak1989}.
For strongly correlated systems, however, the role of the
Berry phase for electronic topology is an open question.

Here we define a Berry curvature in terms of the Green's function eigenfunctions, one for each frequency. A frequency-dependent Chern number follows. It allows us to demonstrate the quantization of the corresponding Berry flux for crossings of the zeros
and, correspondingly, a criterion for its topological nature.
We argue that a computational approach for the frequency-dependent Berry flux facilitates a thorough search for correlated topological materials from 
available databases.
\medskip

\noindent
{\bf Defining 
frequency-dependent Berry curvature from Green's function}

We now define the 
Green's-function-based
Berry curvature. 
   For a given model, consider a Hermitian form of the Green's function, $\mathscr G(\bs k, \omega)$, and its
$\alpha$-th eigenvector $|\phi_{\alpha}(\bs k, \omega)\rangle$ 
   (defined more precisely in the Methods section),
   we can write the eigenvalue equation  
  \beq
  \mathscr G(\bs k, \omega) |\phi_{\alpha}(\bs k, \omega)\rangle = \lambda_{\alpha}(\bs k, \omega) |\phi_{\alpha}(\bs k, \omega)\rangle.
  \eeq 
  We have denoted the $\alpha$-th real eigenvalue of the Green's function $\mathscr G(\bs k, \omega)$ 
  as $\lambda_{\alpha}(\bs k, \omega) $,
  and will refer 
to the ${\bs k}$ dispersion of the eigenvalues as Green's function `bands'.
In terms of these eigenstates, we can define the Berry curvature at momentum $\bs k$ {\it and frequency $\omega$} in terms of
  \beq 
  B_{ij}(\bs k, \omega) \, = \,\sum_{\alpha \in \mathscr S} \Big[ \langle \partial_i \phi_{\alpha}(\bs k, \omega) | \partial_j\phi_{\alpha}(\bs k, \omega)\rangle
  \,-\, \langle \partial_j \phi_{\alpha}(\bs k, \omega) | \partial_i\phi_{\alpha}(\bs k, \omega)\rangle \Big]. \label{Eq:BerryCurvature}
  \eeq
Here $B_{ij}(\bs k, \omega)$ are the components of the Berry curvature vector $B_{l} (\bs k, \omega) = \epsilon_{ijl} B_{ij} (\bs k, \omega)$, with $\epsilon_{ijl}$ being the anti-symmetric tensor; $\alpha \in \mathscr S $ indicates that the sum runs over eigenstates $|\phi_{\alpha}(\bs k, \omega)\rangle$, the  
topological properties of which are of interest. In practice, the subspace $\mathscr S $ corresponds to Green's function eigenvector(s) of a `band' or a subset of (non-crossing) `bands' [\textit{cf}. Fig.\ref{ZeroMonopole}(b, c) where the bands are specified in terms of $\omega - \lambda_{\alpha}(\omega)^{-1}$] at a given $\omega = \omega_0$.  For a given $k_z$ and $\omega$, we can then 
specify a frequency-dependent Chern number in terms of the Berry curvature as 
  \beq
  C(\omega, k_z) = \frac{i}{2\pi}\int d\bs k_{\perp}  B_{xy}(\bs k_{\perp}, k_z, \omega),
  \label{Eq:SpinWinding}
  \eeq
    where $\bs k_{\perp} \equiv  (k_x, k_y)$. 
We note that,
one can also realize 
a topological classification of the Green's functions 
[see Sec.~\ref{app:N-band} of the  Supplemental Information (SI)]
that does not require the notion of ``filled'' eigenstates for defining a topological invariant.
Here, due to our interest in classifying degeneracies of a pair of eigenstates of the Green's function, we have chosen to work with a subset of eigenstates that give rise to such degeneracies and, correspondingly,
a non-trivial set $\mathscr S$.

For definiteness, we will 
consider $G_+(\bs k, \omega)$, which is
the sum of retarded and advanced Green's functions (see Methods for further details), and
determine the eigenvectors $|\phi_{\alpha}(\bs k, \omega)\rangle$, 
of $\mathscr G = G_+(\bs k, \omega)^{-1}$\,.
This quantity
can be readily computed numerically. 
One can alternatively focus on the eigenvectors of $G_+(\bs k, \omega)$.
We note
 that the noninteracting limit
is recovered by
setting all the Coulomb interactions to zero, in which case the Green's function eigenvectors $|\phi_{\alpha}(\bs k, \omega)\rangle$ are simply the Bloch functions (see the Methods).

\medskip

 \noindent
{\bf Model}
\\
To illustrate our  
approach,
we consider a body-centered tetragonal lattice  model with two sub-lattices (see Supplemental Information (SI), Sec.~\ref{app:lattice}, Fig.~\ref{fig:lattice}).
The total Hamiltonian is given by $\mathscr{H} = \mathscr{H}_0 + \mathscr{H}_I$. 
The kinetic part, specified in the Methods,
corresponds to two copies of Weyl semimetals with the same chirality.
It breaks time-reversal symmetry but preserves $SU(2)$ sub-lattice symmetry. 
The degenerate Weyl points are located at $k_z = 
\pm k^*$. To achieve solubility,
we focus on the 
four-fermion interaction terms 
 local in momentum space~\cite{HK1992} in line with Ref.~\cite{Setty2023}.  This is given as
 \beq
\mathscr H_I \,=\,  \frac{\alpha}{2} \sum_{\bs k}  \Phi_{\bs k}^\dagger  \Phi_{\bs k} + \frac{U_c}{2} \sum_{\bs k} \left( \Phi_{\bs k}^\dagger  \Phi_{\bs k}\right)^2 
\,+\,\frac{U_s}{2} \sum_{\bs k} \qty( \Phi_{\bs k}^\dagger \tau_0 \otimes \sigma_3 \Phi_{\bs k})^2.
 \label{Eq:FullHamiltonian}
 \eeq
Here $\alpha$ is a constant shift to the density operator whereas $U_c$ and $U_s$ are the interaction parameters in the total charge and spin sectors respectively. 

In Fig.~\ref{ZeroMonopole}, panels (b,c) show a schematic of the Green's function `bands', 
along high symmetry lines for fixed values of $k_z = k^*$ and $k_z \neq k^*$. 
A spectral crossing 
occurs at $k^*$
[Fig.~\ref{ZeroMonopole}(a)]. Eigenvectors of the lowest pair of these bands are used to determine $B_{ij}(\bs k, \omega)$ in Eq.~\ref{Eq:BerryCurvature}.  
It is instructive to note that setting the interactions to zero reduces $|\phi_{\alpha}(\bs k, \omega) \rangle$ to the filled eigenstates of the noninteracting band structure.

\medskip

\begin{figure}[!t]
    \centering
    \includegraphics[width=1\linewidth]{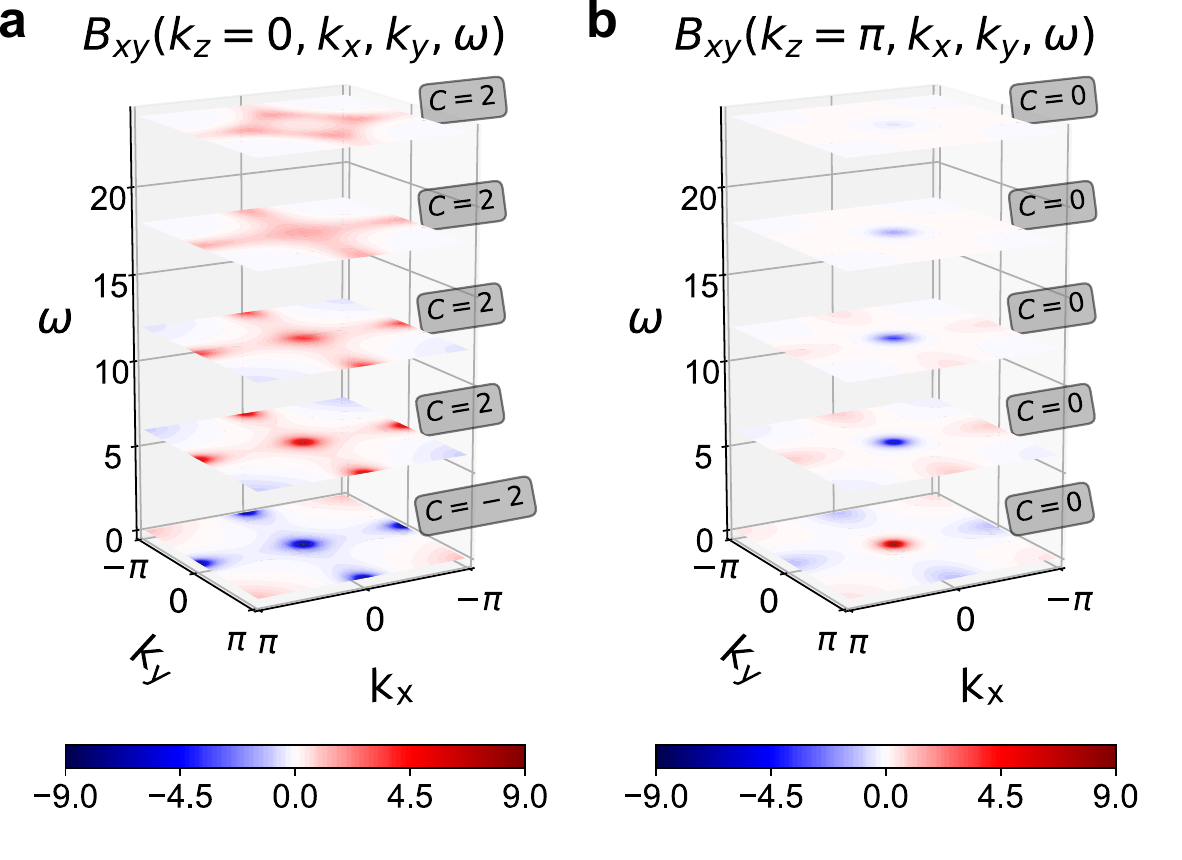}
    \caption{\textbf{Frequency dependent Berry Curvature and Chern number.} 
    Plots of the Berry curvature $B_{ij}(\bs k, \omega)$ as a function of frequency for $k_z=0$ (panel \textbf{(a)}) and $k_z = \pi$ (panel \textbf{(b)}) computed from the eigenvectors of $G_+(\bs k, \omega)^{-1}$.  Red (blue) colors in the intensity color scale denote positive (negative) Berry curvature. The intensity of the color denotes the magnitude of the Berry curvature. The frequency-dependent Chern number for each pair of $\omega, k_z$ is shown as well. Above a critical value of $k_z = k^* = \pi/2$, the interacting system is topologically trivial with $C(\omega, k_z> k^*)=0$. Below the critical value, there is a non-trivial frequency-dependent Chern number obtained from a Berry curvature that varies with frequency and $k_z$. Note that the frequency-dependent Chern number in between the zeros is opposite to that near the poles due to 
    an 
    opposite phase of the kinetic Hamiltonian that appears as a pole in the self-energy. See also Fig.~\ref{Fig:3DBerryCurvature} in SI Sec.~\ref{app:extraplots}. 
    }
    \label{Fig:BerryCurvature-Main}
\end{figure}

 \noindent
{\bf Frequency-dependent Berry curvature
for Green's function 
zeros}
\\
While $G_+(\bs k, \omega)^{-1}$ as well as its eigenvalues and eigenvectors can be determined analytically in simplified limits (see more below), we evaluate here its exact numerical solution to obtain the Berry curvature. In Fig.~\ref{Fig:BandStructureGInv} (SI, Sec.~\ref{app:extraplots}), we provide the eigenvalues of $G_+(\bs k, \omega)^{-1}$ for several pairs of momentum-frequency slices ($k_z, \omega$) across the spectral crossing point $k^*$.  
Fig.~\ref{Fig:BerryCurvature-Main} 
plots the Berry curvature $B_{ij}(\bs k, \omega)$ as a function of frequency computed by numerically exact diagonalization of $\mathscr H$ for each momentum. The $SU(2)$ sub-lattice symmetry ensures that the Berry curvature for the two sublattices is the same. The left (right) panel corresponds to $k_z=0$ ($k_z = \pi$) where the noninteracting system is topologically non-trivial (trivial). Each panel contains planes of constant frequency and the frequency-dependent Chern number evaluated for a given pair of parameters $(\omega, k_z)$. 

The non-trivial variation of the Berry curvature as a function of frequency
is a characteristic of 
an interacting system and is absent in 
the noninteracting case.
We further discuss the non-trivial frequency dependence of the Berry curvature in the SI (Sec.~\ref{app:FDBC}). Additional three dimensional plots of the Berry curvature as a function of $(k_x, k_y, k_z)$ and frequency also appear in the SI (Sec.~\ref{app:extraplots}). 

\medskip
\noindent 
{\bf Berry flux quantization for Green's function zeros}

\begin{figure}[!t]
\centering
\includegraphics[width=0.8\columnwidth]{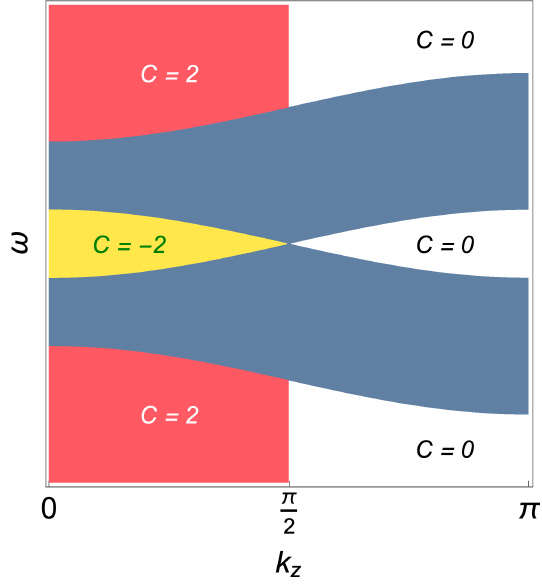}
\caption{\textbf{Frequency dependent Chern number associated with spectral gaps. }Schematic representation of the frequency dependent Chern number ($C$) distribution as a function of frequency and $k_z$ in the vicinity of the bands of zeros. 
The latter locate within the region shaded in blue (color online) where the eigenvectors of $G_+^{-1}$ are not well-defined; consequently, a Chern invariant cannot be associated with them.
Here, 
the electronic bands in the noninteracting limit cross at $\mathbf k = \mathbf k^* \equiv  (0,0,\pi/2)$. 
}
\label{fig:summary}
\end{figure}

A frequency-dependent Chern number can now be calculated for both 
$k_z=0$ and $k_z = \pi$. We find 
it to be quantized for each frequency plan in both cases.
 In the SI (Sec.~\ref{app:N-band}), we provide general arguments for the existence of such a quantization condition.  Analogous to the noninteracting Chern number, the frequency-dependent Chern number picks up a non-zero (zero) integer for $k_z =0< k^*$ ($k_z =\pi > k^*$).  A schematic of the numerically evaluated frequency-dependent Chern numbers for different continuous values of $(\omega, k_z)$ is shown in  Fig.~\ref{fig:summary}.  Below (above) a critical value of $k_z = k^* = \pi/2$, the interacting system is topologically non-trivial (trivial) for all frequency planes. 

Due to the jump 
in the quantized value of the frequency-dependent Chern number across a critical $k_z$ for all frequencies, we can conclude that spectral crossings act as monopoles that source and sink Berry flux (see below and the SI, Sec.~\ref{app:pillbox}) 
even in the case of zeros.
This result is the foundation of our schematic conclusion in Fig.~\ref{ZeroMonopole} (a). 
We return to the generality of this 
  Berry flux quantization 
 from a spectral crossing below (see the Methods). 

 Finally, for a fixed value of $k_z \neq k_z^*$, we find a non-trivial frequency dependence of the quantization value -- if the frequency-dependent Chern number is $+C$ near the upper and lower Hubbard bands (red region in Fig.~\ref{fig:summary}), its value switches to $-C$ in vicinity of the zero surfaces (yellow region in Fig.~\ref{fig:summary}). This is because 
 the hierarchical arrangement of the Green's function `bands' near the poles  can differ from that near the zeros (blue regions in Fig.~\ref{fig:summary}), and can lead to a non-trivial variation of the Berry curvature and frequency-dependent Chern number (see further discussions in the SI, Sec.~\ref{app:FDBC}). 

We close this part with several observations.
 First,
 the frequency dependence of both the Berry curvature and frequency-dependent Chern number is ultimately a correlation driven effect. In the SI (Sec.~\ref{app:commuting}), we discuss a simpler analytically solvable limit where we derive some of the aforementioned conclusions and compare them with the exact solution. 
Second, we reiterate 
that one can alternatively use the eigenvectors of $G_+(\bs k, \omega)$ to characterize interacting topology. We discuss this case in the SI (Sec.~\ref{app:Gplus}). 
Third, our analysis can be readily extended to more general models (see the Methods). 
Fourth, 
our work provides the theoretical foundation to characterize strongly correlated topological semimetals.
In interacting settings, the notion of Dirac or Weyl nodes is replaced by the crossing of the single-particle spectral functions at the nodal wavevector \cite{Hu-Si2021}.

Finally, it has now been established that, in Mott insulators, the Green's function zeros contribute to physical (i.e., measurable) properties~\cite{AGD, sym_green, SettySi2023b}
in a consistent way \cite{SettySi2023b} 
(see the SI,
Sec.~\ref{app:PhysicalProperties}). In particular, by showing that the zeros are accompanied by a quantized Berry phase, our work provides the foundation for the notion that the zeros contribute to quantum oscillations~\cite{Fabrizio2022}.

\medskip

 \noindent
{\bf Topological materials diagnosis}

Our framework sets the stage for developing of a computational
approach
 to search for topology in correlated materials through zeros. 
 These calculations can be carried out by state-of-the-art materials codes such as  
those based on the dynamical mean field theory (DMFT)~\cite{Kotliar2006},
 from which we can obtain the wavevector-dependent self-energy and therefore the Hermitian Green's function.
 Such a computational method can be applied for topological materials search through publicly available databases.
Examples of real materials to explore include Bi$_2$CuO$_4$, in which the bandstructure has been shown to host
high-order (eight-fold degenerate)
nodal points \cite{Bradlyn2016}, and correlations are strong
\cite{DiSante2017,goldoni1994electronic}. One can thus ask whether its paramagnetic
Mott insulating state, above the N\'{e}el temperature ($50$ K) \cite{garcia1990crystal},
is topological.
Another example is the
van der Waals compound
Nb$_3$Cl$_8$, which is a
Mott insulator 
based on  
a half-filled flat band \cite{Gao2022.x}.

\medskip

 \noindent
{\bf Summary and Discussion}

To conclude, we have shown how Mott insulators, arguably the most canonical setting of strong correlation physics, are topological. Central to our approach,
we have advanced a new framework to characterize and diagnose electronic topology in strongly correlated settings. Based on Green's functions, we 
 define a frequency-dependent Berry curvature and Chern number.
 We illustrate the approach through 
 a Mott insulator obtained in a system with doubled Weyl nodal excitations in the underlying electronic structure, 
and demonstrate a quantized monopole Berry charge both from our model calculation and based on general arguments. The result shows that generic symmetry protected spectral crossings can act as sources and sinks of Berry curvature in correlated settings. 
Our framework 
lays the groundwork for future explorations of strongly correlated topological quantum materials, in particular the development of first-principles methods for the search for topological 
Mott insulating materials.

\medskip
\medskip
\medskip

\noindent{\bf\large Methods}
\medskip

 \noindent
{\bf Determining Green's function eigenvectors}
\\
The direct use of the full interacting single-particle 
Green's function eigenvectors is inconvenient;
this is so for the Green's function in Matsubara frequency, as well as for the retarded and advanced Green's functions.
Green's functions by themselves are non-Hermitian; consequently, the relationship between their eigenvectors to frequency-dependent Chern number 
is not straightforward due to the necessity of modified notions of orthonormality and completeness~\cite{Ashida2020, Sato-Review}.

We instead 
consider the
Hermitian combination $\mathscr G(\bs k, \omega) = G_+(\bs k, \omega)^{-1}$.
Here,
$ G_+
\equiv 
G^r
+ G^a
$,
where $G^{r,a}$ are the retarded and advanced Green's functions.
The retarded Green's function 
are defined by
\beq
G^r(\bs k, \omega) = \frac{1}{\omega + i \delta - H_0 - \left(\Sigma'(\bs k, \omega) + i \Sigma''(\bs k, \omega) \right)}
\eeq
and the advanced Green's function follows as the Hermitian conjugate $G^a(\bs k, \omega) = G^r(\bs k, \omega)^{\dagger}$. The real and imaginary parts of the self-energy are denoted by $\Sigma'(\bs k, \omega)$ and $\Sigma''(\bs k, \omega)$ respectively and $H_0$ is the noninteracting Hamiltonian. 
One can readily compute $G_{\pm}(\bs k, \omega) = G^r(\bs k, \omega) \pm G^a(\bs k, \omega)$ and the eigenvectors $|\phi_{\alpha}(\omega, 
 \bs k) \rangle$ of $\mathscr G$.  

\medskip
\noindent
{\bf Specification of the model}
\\
Our model is defined on 
a body-centered tetragonal lattice 
with two sub-lattices
(SI, Sec.~\ref{app:lattice}, Fig.~\ref{fig:lattice}). 
We begin by specifying the 
noninteracting 
part. It is given by
$\mathscr{H}_0 = \sum_{\bs k} \Phi^{\dagger}(\bs k) \left [ \tau_0 \otimes h_0(\bs k) \right ] \Phi(\bs k)$ in the basis $\Phi_{\bs k} = (
c_{A, \Uparrow}, c_{A, \Downarrow}, c_{B, \Uparrow}, c_{B, \Downarrow}
)_{\bs k}^\intercal$. Here, $c_{j, s, \bs k}$ are the destruction operators for an electron in sublattice (or orbital) $j$, spin $s$ and momentum $\bs k$.  $j = A, B$ are the sub-lattice labels, $s= \Uparrow, \Downarrow$ are the physical spins, and $\tau_0$ is the identity matrix in the sublattice (or orbital)  space. We have further defined $h_0(\bs k) = \vec n(\bs k). \vec{\sigma} - \mu$ 
where the vector $\vec n(\bs k) \equiv \{\sin k_x, \sin k_y, \Delta - \sum_{i=x, y, z}\cos k_i \}$, $\vec{\sigma}$ are the Pauli matrices in physical spin space and $\mu$ is the chemical potential~\cite{yang2011quantum, delplace2012, zyuzin2012}. 

The 
model 
corresponds to two copies of Weyl semimetals with the same chirality (as opposed to a Dirac semimetal with an $U(1)$ chiral symmetry; see SI, Sec.~\ref{app:lattice}). The presence of $\tau_0$ in the sub-lattice space means that the kinetic energy does not couple the two sub-lattices (denoted by the red and green circles in Fig.~\ref{fig:lattice}(a) of SI Sec.~\ref{app:lattice}).
It breaks time-reversal symmetry but preserves $SU(2)$ sub-lattice symmetry. The latter ensures that the orbital resolved Berry curvature for the two sublattices is the same.
Our results can be generalized to the case of rotational-symmetry protected Dirac semimetals where $SU(2)$ rotations in spin and sub-lattice are broken in the presence of space-time inversion symmetry~\cite{tyner2023}.
We set the parameter $\Delta =2$ to obtain degenerate Weyl points at $k_z = \pm \frac{\pi}{2} \equiv \pm k^*$. 

 The interaction Hamiltonian is given in Eq.\,\ref{Eq:FullHamiltonian}.
 For generic momenta, the four fermion term $U_c$ commutes with the kinetic energy whereas the $U_s$ term does not. Unlike the kinetic energy, however, both the four fermion terms mix the $A$ and $B$ sublattices.

 \medskip
\noindent
{\bf Berry curvature in the noninteracting limit}
\\
 In the noninteracting limit, the Green's function Berry curvature lacks any non-trivial frequency dependence.
This is as expected,
 given that the noninteracting Hamiltonian trivially commutes with the Green's function. In the 
 specific model we studied, all the  nodes at $k_z = k^*$ of the interacting problem collapse to a single node which is formed by the degeneracy of the upper and lower two bands. 

 \medskip
 \noindent
 {\bf 
 Berry flux quantization in more general models}
 \\
 In the
  specific calculation we described in the main text, we focused on a four band model of doubled Weyl-fermions with a non-trivial frequency-dependent Chern number. However, our arguments are generic to any number of bands.
We studied a doubled Weyl semimetal with an $SU(2)$ sublattice-rotational symmetry that is preserved in the Mott insulating state. 
The sublattice-rotational symmetry constrains the interacting Green's function to be block diagonal, with each block mapping the $(k_x, k_y)$ planes to $\frac{SU(2)}{U(1)} \equiv S^2$, the two-sphere, at fixed $\omega$ and $k_z$.
Such maps can be classified by the second homotopy group of $S^2$, with 
$\pi_2\left(S^2 \right) = \mathbbm{Z}$.
The frequency-dependent Chern number is a measure of this homotopy invariant. 

In analogy to $N$-band Chern insulators, classification of the eigenvectors of block-diagonal Green's functions can be extended to generic $2N$ band models (see the SI, Sec.~\ref{app:N-band}).
We leave details of such considerations to future works.

We note that our formalism can be applied when the sublattice-$SU(2)$ symmetry is broken down to $U(1)$, and, 
it is possible to define a quantized sublattice-Chern number at fixed $(k_z, \omega)$ (see the SI, Sec.~\ref{app:N-band}, for more details).

Finally, the primary focus of our work has been on the utility of eigenvectors of $G_+(\bs k, \omega)$. For generic interaction problems, 
it is possible that the eigenvectors of both $G_{\pm}(\bs k, \omega)$ are needed.

\medskip
\noindent
{\bf Data availability}\\
 The data that support the findings of this study are available from the corresponding author
upon reasonable request.
\\ \newline
\noindent $^\dagger$ These authors contributed equally.

\noindent
\bibliographystyle{naturemagallauthors}

\bibliography{Zeros.bib}

\begin{thebibliography}{10}
\expandafter\ifx\csname url\endcsname\relax
  \def\url#1{\texttt{#1}}\fi
\expandafter\ifx\csname urlprefix\endcsname\relax\def\urlprefix{URL }\fi
\providecommand{\bibinfo}[2]{#2}
\providecommand{\eprint}[2][]{\url{#2}}

\bibitem{Stormer99}
\bibinfo{author}{Stormer, H.~L.}, \bibinfo{author}{Tsui, D.~C.} \&
  \bibinfo{author}{Gossard, A.~C.}
\newblock \bibinfo{title}{The fractional quantum hall effect}.
\newblock \emph{\bibinfo{journal}{Rev. Mod. Phys.}}
  \textbf{\bibinfo{volume}{71}}, \bibinfo{pages}{S298} (\bibinfo{year}{1999}).

\bibitem{Xie2021}
\bibinfo{author}{Xie, Y.}, \bibinfo{author}{Pierce, A.~T.},
  \bibinfo{author}{Park, J.~M.}, \bibinfo{author}{Parker, D.~E.},
  \bibinfo{author}{Khalaf, E.}, \bibinfo{author}{Ledwith, P.},
  \bibinfo{author}{Cao, Y.}, \bibinfo{author}{Lee, S.~H.},
  \bibinfo{author}{Chen, S.}, \bibinfo{author}{Forrester, P.~R.},
  \bibinfo{author}{Watanabe, K.}, \bibinfo{author}{Taniguchi, T.},
  \bibinfo{author}{Vishwanath, A.}, \bibinfo{author}{Jarillo-Herrero, P.} \&
  \bibinfo{author}{Yacoby, A.}
\newblock \bibinfo{title}{{Fractional Chern insulators in magic-angle twisted
  bilayer graphene}}.
\newblock \emph{\bibinfo{journal}{{Nature}}} \textbf{\bibinfo{volume}{600}},
  \bibinfo{pages}{439--443} (\bibinfo{year}{2021}).

\bibitem{Xu2023}
\bibinfo{author}{Park, H.}, \bibinfo{author}{Cai, J.},
  \bibinfo{author}{Anderson, E.}, \bibinfo{author}{Zhang, Y.},
  \bibinfo{author}{Zhu, J.}, \bibinfo{author}{Liu, X.}, \bibinfo{author}{Wang,
  C.}, \bibinfo{author}{Holtzmann, W.}, \bibinfo{author}{Hu, C.},
  \bibinfo{author}{Liu, Z.}, \bibinfo{author}{Taniguchi, T.},
  \bibinfo{author}{Watanabe, K.}, \bibinfo{author}{Chu, J.-H.},
  \bibinfo{author}{Cao, T.}, \bibinfo{author}{Fu, L.}, \bibinfo{author}{Yao,
  W.}, \bibinfo{author}{Chang, C.-Z.}, \bibinfo{author}{Cobden, D.},
  \bibinfo{author}{Xiao, D.} \& \bibinfo{author}{Xu, X.}
\newblock \bibinfo{title}{Observation of fractionally quantized anomalous hall
  effect}.
\newblock \emph{\bibinfo{journal}{Nature}} \textbf{\bibinfo{volume}{622}},
  \bibinfo{pages}{74--79} (\bibinfo{year}{2023}).

\bibitem{Zeng-fci2023}
\bibinfo{author}{Zeng, Y.}, \bibinfo{author}{Xia, Z.}, \bibinfo{author}{Kang,
  K.}, \bibinfo{author}{Zhu, J.}, \bibinfo{author}{Kn\"uppel, P.},
  \bibinfo{author}{Vaswani, C.}, \bibinfo{author}{Watanabe, K.},
  \bibinfo{author}{Taniguchi, T.}, \bibinfo{author}{Mak, K.~F.} \&
  \bibinfo{author}{Shan, J.}
\newblock \bibinfo{title}{{Thermodynamic evidence of fractional Chern insulator
  in moir\'e MoTe$_2$}}.
\newblock \emph{\bibinfo{journal}{{Nature}}} \textbf{\bibinfo{volume}{622}},
  \bibinfo{pages}{69--73} (\bibinfo{year}{2023}).

\bibitem{Lai2018}
\bibinfo{author}{Lai, H.-H.}, \bibinfo{author}{Grefe, S.~E.},
  \bibinfo{author}{Paschen, S.} \& \bibinfo{author}{Si, Q.}
\newblock \bibinfo{title}{Weyl-{Kondo} semimetal in heavy-fermion systems}.
\newblock \emph{\bibinfo{journal}{Proc. Natl. Acad. Sci. U.S.A.}}
  \textbf{\bibinfo{volume}{115}}, \bibinfo{pages}{93} (\bibinfo{year}{2018}).

\bibitem{Chen-Si2022}
\bibinfo{author}{Chen, L.}, \bibinfo{author}{Setty, C.}, \bibinfo{author}{Hu,
  H.}, \bibinfo{author}{Vergniory, M.~G.}, \bibinfo{author}{Grefe, S.~E.},
  \bibinfo{author}{Fischer, L.}, \bibinfo{author}{Yan, X.},
  \bibinfo{author}{Eguchi, G.}, \bibinfo{author}{Prokofiev, A.},
  \bibinfo{author}{Paschen, S.}, \bibinfo{author}{Cano, J.} \&
  \bibinfo{author}{Si, Q.}
\newblock \bibinfo{title}{Topological semimetal driven by strong correlations
  and crystalline symmetry}.
\newblock \emph{\bibinfo{journal}{Nat. Phys.}} \textbf{\bibinfo{volume}{18}},
  \bibinfo{pages}{1341} (\bibinfo{year}{2022}).

\bibitem{Dzsaber2017}
\bibinfo{author}{Dzsaber, S.}, \bibinfo{author}{Prochaska, L.},
  \bibinfo{author}{Sidorenko, A.}, \bibinfo{author}{Eguchi, G.},
  \bibinfo{author}{Svagera, R.}, \bibinfo{author}{Waas, M.},
  \bibinfo{author}{Prokofiev, A.}, \bibinfo{author}{Si, Q.} \&
  \bibinfo{author}{Paschen, S.}
\newblock \bibinfo{title}{Kondo insulator to semimetal transformation tuned by
  spin-orbit coupling}.
\newblock \emph{\bibinfo{journal}{Phys. Rev. Lett.}}
  \textbf{\bibinfo{volume}{118}}, \bibinfo{pages}{246601}
  (\bibinfo{year}{2017}).

\bibitem{Dzs-giant21.1}
\bibinfo{author}{Dzsaber, S.}, \bibinfo{author}{Yan, X.},
  \bibinfo{author}{Taupin, M.}, \bibinfo{author}{Eguchi, G.},
  \bibinfo{author}{Prokofiev, A.}, \bibinfo{author}{Shiroka, T.},
  \bibinfo{author}{Blaha, P.}, \bibinfo{author}{Rubel, O.},
  \bibinfo{author}{Grefe, S.~E.}, \bibinfo{author}{Lai, H.-H.},
  \bibinfo{author}{Si, Q.} \& \bibinfo{author}{Paschen, S.}
\newblock \bibinfo{title}{Giant spontaneous hall effect in a nonmagnetic {W}eyl
  {K}ondo semimetal}.
\newblock \emph{\bibinfo{journal}{Proc. Natl. Acad. Sci. U.S.A.}}
  \textbf{\bibinfo{volume}{118}}, \bibinfo{pages}{e2013386118}
  (\bibinfo{year}{2021}).

\bibitem{abrikosov2012}
\bibinfo{author}{Abrikosov, A.~A.}, \bibinfo{author}{Gorkov, L.~P.} \&
  \bibinfo{author}{Dzyaloshinski, I.~E.}
\newblock \emph{\bibinfo{title}{Methods of quantum field theory in statistical
  physics}} (\bibinfo{publisher}{Courier Corporation}, \bibinfo{year}{2012}).

\bibitem{Armitage2017}
\bibinfo{author}{Armitage, N.~P.}, \bibinfo{author}{Mele, E.~J.} \&
  \bibinfo{author}{Vishwanath, A.}
\newblock \bibinfo{title}{{Weyl} and {Dirac} semimetals in three-dimensional
  solids}.
\newblock \emph{\bibinfo{journal}{Rev. Mod. Phys.}}
  \textbf{\bibinfo{volume}{90}}, \bibinfo{pages}{015001}
  (\bibinfo{year}{2018}).

\bibitem{Nagaosa2020}
\bibinfo{author}{Nagaosa, N.}, \bibinfo{author}{Morimoto, T.} \&
  \bibinfo{author}{Tokura, Y.}
\newblock \bibinfo{title}{Transport, magnetic and optical properties of weyl
  materials}.
\newblock \emph{\bibinfo{journal}{Nat. Rev. Mater.}}
  \textbf{\bibinfo{volume}{5}}, \bibinfo{pages}{621} (\bibinfo{year}{2020}).

\bibitem{Bradlyn2017}
\bibinfo{author}{Bradlyn, B.}, \bibinfo{author}{Elcoro, L.},
  \bibinfo{author}{Cano, J.}, \bibinfo{author}{Vergniory, M.~G.},
  \bibinfo{author}{Wang, Z.}, \bibinfo{author}{Felser, C.},
  \bibinfo{author}{Aroyo, M.~I.} \& \bibinfo{author}{Bernevig, B.~A.}
\newblock \bibinfo{title}{Topological quantum chemistry}.
\newblock \emph{\bibinfo{journal}{Nature}} \textbf{\bibinfo{volume}{547}},
  \bibinfo{pages}{298} (\bibinfo{year}{2017}).

\bibitem{Cano2018}
\bibinfo{author}{Cano, J.}, \bibinfo{author}{Bradlyn, B.},
  \bibinfo{author}{Wang, Z.}, \bibinfo{author}{Elcoro, L.},
  \bibinfo{author}{Vergniory, M.~G.}, \bibinfo{author}{Felser, C.},
  \bibinfo{author}{Aroyo, M.~I.} \& \bibinfo{author}{Bernevig, B.~A.}
\newblock \bibinfo{title}{Building blocks of topological quantum chemistry:
  Elementary band representations}.
\newblock \emph{\bibinfo{journal}{Phys. Rev. B}} \textbf{\bibinfo{volume}{97}},
  \bibinfo{pages}{035139} (\bibinfo{year}{2018}).

\bibitem{Po2017}
\bibinfo{author}{Po, H.~C.}, \bibinfo{author}{Vishwanath, A.} \&
  \bibinfo{author}{Watanabe, H.}
\newblock \bibinfo{title}{Symmetry-based indicators of band topology in the 230
  space groups}.
\newblock \emph{\bibinfo{journal}{Nat. Commun.}} \textbf{\bibinfo{volume}{8}},
  \bibinfo{pages}{50} (\bibinfo{year}{2017}).

\bibitem{Watanabe2017}
\bibinfo{author}{Watanabe, H.}, \bibinfo{author}{Po, H.~C.},
  \bibinfo{author}{Zaletel, M.~P.} \& \bibinfo{author}{Vishwanath, A.}
\newblock \bibinfo{title}{Filling-enforced gaplessness in band structures of
  the 230 space groups}.
\newblock \emph{\bibinfo{journal}{Phys. Rev. Lett.}}
  \textbf{\bibinfo{volume}{117}}, \bibinfo{pages}{096404}
  (\bibinfo{year}{2016}).

\bibitem{cano2021band}
\bibinfo{author}{Cano, J.} \& \bibinfo{author}{Bradlyn, B.}
\newblock \bibinfo{title}{Band representations and topological quantum
  chemistry}.
\newblock \emph{\bibinfo{journal}{Annu.\ Rev.\ Condens.\ Matter Phys.}}
  \textbf{\bibinfo{volume}{12}}, \bibinfo{pages}{225} (\bibinfo{year}{2021}).

\bibitem{Wang-Zhang_PRX2012}
\bibinfo{author}{Wang, Z.} \& \bibinfo{author}{Zhang, S.-C.}
\newblock \bibinfo{title}{Simplified topological invariants for interacting
  insulators}.
\newblock \emph{\bibinfo{journal}{Phys. Rev. X}} \textbf{\bibinfo{volume}{2}},
  \bibinfo{pages}{031008} (\bibinfo{year}{2012}).

\bibitem{Wang-Yan2013}
\bibinfo{author}{Wang, Z.} \& \bibinfo{author}{Yan, B.}
\newblock \bibinfo{title}{Topological hamiltonian as an exact tool for
  topological invariants}.
\newblock \emph{\bibinfo{journal}{J. Phys-Condens. Mat.}}
  \textbf{\bibinfo{volume}{25}}, \bibinfo{pages}{155601}
  (\bibinfo{year}{2013}).

\bibitem{Iraola21}
\bibinfo{author}{Iraola, M.}, \bibinfo{author}{Heinsdorf, N.},
  \bibinfo{author}{Tiwari, A.}, \bibinfo{author}{Lessnich, D.},
  \bibinfo{author}{Mertz, T.}, \bibinfo{author}{Ferrari, F.},
  \bibinfo{author}{Fischer, M.~H.}, \bibinfo{author}{Winter, S.~M.},
  \bibinfo{author}{Pollmann, F.}, \bibinfo{author}{Neupert, T.},
  \bibinfo{author}{Valent\'{\i}, R.} \& \bibinfo{author}{Vergniory, M.~G.}
\newblock \bibinfo{title}{Towards a topological quantum chemistry description
  of correlated systems: The case of the hubbard diamond chain}.
\newblock \emph{\bibinfo{journal}{Phys. Rev. B}}
  \textbf{\bibinfo{volume}{104}}, \bibinfo{pages}{195125}
  (\bibinfo{year}{2021}).

\bibitem{Lessnich2021}
\bibinfo{author}{Lessnich, D.}, \bibinfo{author}{Winter, S.~M.},
  \bibinfo{author}{Iraola, M.}, \bibinfo{author}{Vergniory, M.~G.} \&
  \bibinfo{author}{Valent\'{\i}, R.}
\newblock \bibinfo{title}{Elementary band representations for the
  single-particle green's function of interacting topological insulators}.
\newblock \emph{\bibinfo{journal}{Phys. Rev. B}}
  \textbf{\bibinfo{volume}{104}}, \bibinfo{pages}{085116}
  (\bibinfo{year}{2021}).

\bibitem{Soldini2022}
\bibinfo{author}{Soldini, M.~O.}, \bibinfo{author}{Astrakhantsev, N.},
  \bibinfo{author}{Iraola, M.}, \bibinfo{author}{Tiwari, A.},
  \bibinfo{author}{Fischer, M.~H.}, \bibinfo{author}{Valent{\'\i}, R.},
  \bibinfo{author}{Vergniory, M.~G.}, \bibinfo{author}{Wagner, G.} \&
  \bibinfo{author}{Neupert, T.}
\newblock \bibinfo{title}{Interacting topological quantum chemistry of mott
  atomic limits}.
\newblock \emph{\bibinfo{journal}{arXiv preprint arXiv:2209.10556}}
  (\bibinfo{year}{2022}).

\bibitem{Hu-Si2021}
\bibinfo{author}{Hu, H.}, \bibinfo{author}{Chen, L.}, \bibinfo{author}{Setty,
  C.}, \bibinfo{author}{Grefe, S.~E.}, \bibinfo{author}{Prokofiev, A.},
  \bibinfo{author}{Kirchner, S.}, \bibinfo{author}{Paschen, S.},
  \bibinfo{author}{Cano, J.} \& \bibinfo{author}{Si, Q.}
\newblock \bibinfo{title}{Topological semimetals without quasiparticles}.
\newblock \emph{\bibinfo{journal}{arXiv preprint arXiv:2110.06182}}
  (\bibinfo{year}{2021}).

\bibitem{Setty2023}
\bibinfo{author}{Setty, C.}, \bibinfo{author}{Sur, S.}, \bibinfo{author}{Chen,
  L.}, \bibinfo{author}{Xie, F.}, \bibinfo{author}{Hu, H.},
  \bibinfo{author}{Paschen, S.}, \bibinfo{author}{Cano, J.} \&
  \bibinfo{author}{Si, Q.}
\newblock \bibinfo{title}{Symmetry constraints and spectral crossing in a mott
  insulator with green's function zeros}.
\newblock \emph{\bibinfo{journal}{arXiv preprint arXiv:2301.13870}}
  (\bibinfo{year}{2023}).

\bibitem{AGD}
\bibinfo{author}{Abrikosov, A.~A.}, \bibinfo{author}{Gorkov, L.~P.} \&
  \bibinfo{author}{Dzyaloshinski, I.~E.}
\newblock \emph{\bibinfo{title}{Methods of quantum field theory in statistical
  physics}} (\bibinfo{publisher}{Courier Corporation}, \bibinfo{year}{2012}).

\bibitem{Dzyaloshinskii2003}
\bibinfo{author}{Dzyaloshinskii, I.}
\newblock \bibinfo{title}{Some consequences of the luttinger theorem: The
  luttinger surfaces in non-fermi liquids and mott insulators}.
\newblock \emph{\bibinfo{journal}{Phys. Rev. B}} \textbf{\bibinfo{volume}{68}},
  \bibinfo{pages}{085113} (\bibinfo{year}{2003}).

\bibitem{sym_green}
\bibinfo{author}{Gurarie, V.}
\newblock \bibinfo{title}{Single-particle green's functions and interacting
  topological insulators}.
\newblock \emph{\bibinfo{journal}{Phys. Rev. B}} \textbf{\bibinfo{volume}{83}},
  \bibinfo{pages}{085426} (\bibinfo{year}{2011}).

\bibitem{Gurarie2011-2}
\bibinfo{author}{Essin, A.~M.} \& \bibinfo{author}{Gurarie, V.}
\newblock \bibinfo{title}{Bulk-boundary correspondence of topological
  insulators from their respective green's functions}.
\newblock \emph{\bibinfo{journal}{Phys. Rev. B}} \textbf{\bibinfo{volume}{84}},
  \bibinfo{pages}{125132} (\bibinfo{year}{2011}).

\bibitem{Yunoki2017}
\bibinfo{author}{Seki, K.} \& \bibinfo{author}{Yunoki, S.}
\newblock \bibinfo{title}{Topological interpretation of the luttinger theorem}.
\newblock \emph{\bibinfo{journal}{Phys. Rev. B}} \textbf{\bibinfo{volume}{96}},
  \bibinfo{pages}{085124} (\bibinfo{year}{2017}).

\bibitem{SettySi2023b}
\bibinfo{author}{Setty, C.}, \bibinfo{author}{Xie, F.}, \bibinfo{author}{Sur,
  S.}, \bibinfo{author}{Chen, L.}, \bibinfo{author}{Vergniory, M.~G.} \&
  \bibinfo{author}{Si, Q.}
\newblock \bibinfo{title}{Electronic properties, correlated topology and
  green's function zeros}.
\newblock \emph{\bibinfo{journal}{arXiv preprint arXiv:2309.14340}}
  (\bibinfo{year}{2023}).

\bibitem{wagner2023mott}
\bibinfo{author}{Wagner, N.}, \bibinfo{author}{Crippa, L.},
  \bibinfo{author}{Amaricci, A.}, \bibinfo{author}{Hansmann, P.},
  \bibinfo{author}{Klett, M.}, \bibinfo{author}{K\"onig, E.},
  \bibinfo{author}{Sch\"afer, T.}, \bibinfo{author}{Di~Sante, D.},
  \bibinfo{author}{Cano, J.}, \bibinfo{author}{Millis, A.},
  \bibinfo{author}{Georges, A.} \& \bibinfo{author}{Sangiovanni, G.}
\newblock \bibinfo{title}{Mott insulators with boundary zeros}.
\newblock \emph{\bibinfo{journal}{arXiv preprint arXiv:2301.05588}}
  (\bibinfo{year}{2023}).

\bibitem{Fabrizio2022}
\bibinfo{author}{Fabrizio, M.}
\newblock \bibinfo{title}{Emergent quasiparticles at luttinger surfaces}.
\newblock \emph{\bibinfo{journal}{Nat. Commun.}} \textbf{\bibinfo{volume}{13}},
  \bibinfo{pages}{1} (\bibinfo{year}{2022}).

\bibitem{Fabrizio2023}
\bibinfo{author}{Blason, A.} \& \bibinfo{author}{Fabrizio, M.}
\newblock \bibinfo{title}{Unified role of green's function poles and zeros in
  topological insulators}.
\newblock \emph{\bibinfo{journal}{arXiv preprint arXiv:2304.08180}}
  (\bibinfo{year}{2023}).

\bibitem{Raghu-Zhang2008}
\bibinfo{author}{Raghu, S.}, \bibinfo{author}{Qi, X.-L.},
  \bibinfo{author}{Honerkamp, C.} \& \bibinfo{author}{Zhang, S.-C.}
\newblock \bibinfo{title}{Topological mott insulators}.
\newblock \emph{\bibinfo{journal}{Phys. Rev. Lett.}}
  \textbf{\bibinfo{volume}{100}}, \bibinfo{pages}{156401}
  (\bibinfo{year}{2008}).

\bibitem{Nagaosa2016}
\bibinfo{author}{Morimoto, T.} \& \bibinfo{author}{Nagaosa, N.}
\newblock \bibinfo{title}{Weyl mott insulator}.
\newblock \emph{\bibinfo{journal}{Sci. Rep.}} \textbf{\bibinfo{volume}{6}},
  \bibinfo{pages}{19853} (\bibinfo{year}{2016}).

\bibitem{Phillips2023-NatCommun}
\bibinfo{author}{Mai, P.}, \bibinfo{author}{Zhao, J.},
  \bibinfo{author}{Feldman, B.~E.} \& \bibinfo{author}{Phillips, P.~W.}
\newblock \bibinfo{title}{1/4 is the new 1/2 when topology is intertwined with
  mottness}.
\newblock \emph{\bibinfo{journal}{Nature Communications}}
  \textbf{\bibinfo{volume}{14}}, \bibinfo{pages}{5999} (\bibinfo{year}{2023}).

\bibitem{Pancharatnam1956}
\bibinfo{author}{Pancharatnam, S.}
\newblock \bibinfo{title}{Generalized theory of interference, and its
  applications: Part i. coherent pencils}.
\newblock In \emph{\bibinfo{booktitle}{Proceedings of the Indian Academy of
  Sciences-Section A}}, vol.~\bibinfo{volume}{44}, \bibinfo{pages}{247--262}
  (\bibinfo{organization}{Springer}, \bibinfo{year}{1956}).

\bibitem{Berry1984}
\bibinfo{author}{Berry, M.~V.}
\newblock \bibinfo{title}{Quantal phase factors accompanying adiabatic
  changes}.
\newblock \emph{\bibinfo{journal}{Proceedings of the Royal Society of London.
  A. Mathematical and Physical Sciences}} \textbf{\bibinfo{volume}{392}},
  \bibinfo{pages}{45--57} (\bibinfo{year}{1984}).

\bibitem{Zak1989}
\bibinfo{author}{Zak, J.}
\newblock \bibinfo{title}{Berry's phase for energy bands in solids}.
\newblock \emph{\bibinfo{journal}{Phys. Rev. Lett.}}
  \textbf{\bibinfo{volume}{62}}, \bibinfo{pages}{2747} (\bibinfo{year}{1989}).

\bibitem{HK1992}
\bibinfo{author}{Hatsugai, Y.} \& \bibinfo{author}{Kohmoto, M.}
\newblock \bibinfo{title}{Exactly solvable model of correlated lattice
  electrons in any dimensions}.
\newblock \emph{\bibinfo{journal}{J. Phys. Soc. Jpn.}}
  \textbf{\bibinfo{volume}{61}}, \bibinfo{pages}{2056} (\bibinfo{year}{1992}).

\bibitem{Kotliar2006}
\bibinfo{author}{Kotliar, G.}, \bibinfo{author}{Savrasov, S.~Y.},
  \bibinfo{author}{Haule, K.}, \bibinfo{author}{Oudovenko, V.~S.},
  \bibinfo{author}{Parcollet, O.} \& \bibinfo{author}{Marianetti, C.~A.}
\newblock \bibinfo{title}{Electronic structure calculations with dynamical
  mean-field theory}.
\newblock \emph{\bibinfo{journal}{Rev. Mod. Phys.}}
  \textbf{\bibinfo{volume}{78}}, \bibinfo{pages}{865--951}
  (\bibinfo{year}{2006}).

\bibitem{Bradlyn2016}
\bibinfo{author}{Bradlyn, B.}, \bibinfo{author}{Cano, J.},
  \bibinfo{author}{Wang, Z.}, \bibinfo{author}{Vergniory, M.~G.},
  \bibinfo{author}{Felser, C.}, \bibinfo{author}{Cava, R.~J.} \&
  \bibinfo{author}{Bernevig, B.~A.}
\newblock \bibinfo{title}{Beyond dirac and weyl fermions: Unconventional
  quasiparticles in conventional crystals}.
\newblock \emph{\bibinfo{journal}{Science}} \textbf{\bibinfo{volume}{353}},
  \bibinfo{pages}{aaf5037} (\bibinfo{year}{2016}).

\bibitem{DiSante2017}
\bibinfo{author}{Di~Sante, D.}, \bibinfo{author}{Hausoel, A.},
  \bibinfo{author}{Barone, P.}, \bibinfo{author}{Tomczak, J.~M.},
  \bibinfo{author}{Sangiovanni, G.} \& \bibinfo{author}{Thomale, R.}
\newblock \bibinfo{title}{Realizing double dirac particles in the presence of
  electronic interactions}.
\newblock \emph{\bibinfo{journal}{Phys. Rev. B}} \textbf{\bibinfo{volume}{96}},
  \bibinfo{pages}{121106} (\bibinfo{year}{2017}).

\bibitem{goldoni1994electronic}
\bibinfo{author}{Goldoni, A.}, \bibinfo{author}{del Pennino, U.},
  \bibinfo{author}{Parmigiani, F.}, \bibinfo{author}{Sangaletti, L.} \&
  \bibinfo{author}{Revcolevschi, A.}
\newblock \bibinfo{title}{Electronic structure of {$\rm Bi_2 Cu O_4$}}.
\newblock \emph{\bibinfo{journal}{Phys. Rev. B}} \textbf{\bibinfo{volume}{50}},
  \bibinfo{pages}{10435} (\bibinfo{year}{1994}).

\bibitem{garcia1990crystal}
\bibinfo{author}{Garcia-Munoz, J.}, \bibinfo{author}{Rodriguez-Carvajal, J.},
  \bibinfo{author}{Sapina, F.}, \bibinfo{author}{Sanchis, M.},
  \bibinfo{author}{Ibanez, R.} \& \bibinfo{author}{Beltran-Porter, D.}
\newblock \bibinfo{title}{Crystal and magnetic structures of bi2cuo4}.
\newblock \emph{\bibinfo{journal}{Journal of Physics: Condensed Matter}}
  \textbf{\bibinfo{volume}{2}}, \bibinfo{pages}{2205} (\bibinfo{year}{1990}).

\bibitem{Gao2022.x}
\bibinfo{author}{{Gao}, S.}, \bibinfo{author}{{Zhang}, S.},
  \bibinfo{author}{{Wang}, C.}, \bibinfo{author}{{Tao}, W.},
  \bibinfo{author}{{Liu}, J.}, \bibinfo{author}{{Wang}, T.},
  \bibinfo{author}{{Yuan}, S.}, \bibinfo{author}{{Qu}, G.},
  \bibinfo{author}{{Pan}, M.}, \bibinfo{author}{{Peng}, S.},
  \bibinfo{author}{{Hu}, Y.}, \bibinfo{author}{{Li}, H.},
  \bibinfo{author}{{Huang}, Y.}, \bibinfo{author}{{Zhou}, H.},
  \bibinfo{author}{{Meng}, S.}, \bibinfo{author}{{Yang}, L.},
  \bibinfo{author}{{Wang}, Z.}, \bibinfo{author}{{Yao}, Y.},
  \bibinfo{author}{{Chen}, Z.}, \bibinfo{author}{{Shi}, M.},
  \bibinfo{author}{{Ding}, H.}, \bibinfo{author}{{Jiang}, K.},
  \bibinfo{author}{{Li}, Y.}, \bibinfo{author}{{Shi}, Y.},
  \bibinfo{author}{{Weng}, H.} \& \bibinfo{author}{{Qian}, T.}
\newblock \bibinfo{title}{Mott insulator state in a van der waals flat-band
  compound}.
\newblock \emph{\bibinfo{journal}{arXiv preprint arXiv:2205.11462}}
  (\bibinfo{year}{2022}).

\bibitem{Ashida2020}
\bibinfo{author}{Ashida, Y.}, \bibinfo{author}{Gong, Z.} \&
  \bibinfo{author}{Ueda, M.}
\newblock \bibinfo{title}{Non-hermitian physics}.
\newblock \emph{\bibinfo{journal}{Advances in Physics}}
  \textbf{\bibinfo{volume}{69}}, \bibinfo{pages}{249} (\bibinfo{year}{2020}).

\bibitem{Sato-Review}
\bibinfo{author}{Okuma, N.} \& \bibinfo{author}{Sato, M.}
\newblock \bibinfo{title}{Non-hermitian topological phenomena: A review}.
\newblock \emph{\bibinfo{journal}{Annu. Rev. Condens. Matter Phys.}}
  \textbf{\bibinfo{volume}{14}}, \bibinfo{pages}{83} (\bibinfo{year}{2023}).

\bibitem{yang2011quantum}
\bibinfo{author}{Yang, K.-Y.}, \bibinfo{author}{Lu, Y.-M.} \&
  \bibinfo{author}{Ran, Y.}
\newblock \bibinfo{title}{Quantum hall effects in a weyl semimetal: Possible
  application in pyrochlore iridates}.
\newblock \emph{\bibinfo{journal}{Phys. Rev. B}} \textbf{\bibinfo{volume}{84}},
  \bibinfo{pages}{075129} (\bibinfo{year}{2011}).

\bibitem{delplace2012}
\bibinfo{author}{Delplace, P.}, \bibinfo{author}{Li, J.} \&
  \bibinfo{author}{Carpentier, D.}
\newblock \bibinfo{title}{Topological weyl semi-metal from a lattice model}.
\newblock \emph{\bibinfo{journal}{Europhys. Lett.}}
  \textbf{\bibinfo{volume}{97}}, \bibinfo{pages}{67004} (\bibinfo{year}{2012}).

\bibitem{zyuzin2012}
\bibinfo{author}{Zyuzin, A.}, \bibinfo{author}{Wu, S.} \&
  \bibinfo{author}{Burkov, A.}
\newblock \bibinfo{title}{Weyl semimetal with broken time reversal and
  inversion symmetries}.
\newblock \emph{\bibinfo{journal}{Phys. Rev. B}} \textbf{\bibinfo{volume}{85}},
  \bibinfo{pages}{165110} (\bibinfo{year}{2012}).

\bibitem{tyner2023}
\bibinfo{author}{Tyner, A.~C.}, \bibinfo{author}{Sur, S.},
  \bibinfo{author}{Puggioni, D.}, \bibinfo{author}{Rondinelli, J.~M.} \&
  \bibinfo{author}{Goswami, P.}
\newblock \bibinfo{title}{Topology of three-dimensional dirac semimetals and
  quantum spin hall systems without gapless edge modes}.
\newblock \emph{\bibinfo{journal}{Phys. Rev. Res.}}
  \textbf{\bibinfo{volume}{5}}, \bibinfo{pages}{L012019}
  (\bibinfo{year}{2023}).

\bibitem{wan2011}
\bibinfo{author}{Wan, X.}, \bibinfo{author}{Turner, A.~M.},
  \bibinfo{author}{Vishwanath, A.} \& \bibinfo{author}{Savrasov, S.~Y.}
\newblock \bibinfo{title}{Topological semimetal and fermi-arc surface states in
  the electronic structure of pyrochlore iridates}.
\newblock \emph{\bibinfo{journal}{Phys. Rev. B}} \textbf{\bibinfo{volume}{83}},
  \bibinfo{pages}{205101} (\bibinfo{year}{2011}).

\bibitem{bernevig2006}
\bibinfo{author}{Bernevig, B.~A.}, \bibinfo{author}{Hughes, T.~L.} \&
  \bibinfo{author}{Zhang, S.-C.}
\newblock \bibinfo{title}{Quantum spin hall effect and topological phase
  transition in hgte quantum wells}.
\newblock \emph{\bibinfo{journal}{Science}} \textbf{\bibinfo{volume}{314}},
  \bibinfo{pages}{1757} (\bibinfo{year}{2006}).

\bibitem{yang2014classification}
\bibinfo{author}{Yang, B.-J.} \& \bibinfo{author}{Nagaosa, N.}
\newblock \bibinfo{title}{Classification of stable three-dimensional dirac
  semimetals with nontrivial topology}.
\newblock \emph{\bibinfo{journal}{Nat. Commun.}} \textbf{\bibinfo{volume}{5}},
  \bibinfo{pages}{4898} (\bibinfo{year}{2014}).

\bibitem{avron1983}
\bibinfo{author}{Avron, J.~E.}, \bibinfo{author}{Seiler, R.} \&
  \bibinfo{author}{Simon, B.}
\newblock \bibinfo{title}{Homotopy and quantization in condensed matter
  physics}.
\newblock \emph{\bibinfo{journal}{Phys. Rev. Lett.}}
  \textbf{\bibinfo{volume}{51}}, \bibinfo{pages}{51} (\bibinfo{year}{1983}).

\bibitem{moore2008}
\bibinfo{author}{Moore, J.~E.}, \bibinfo{author}{Ran, Y.} \&
  \bibinfo{author}{Wen, X.-G.}
\newblock \bibinfo{title}{Topological surface states in three-dimensional
  magnetic insulators}.
\newblock \emph{\bibinfo{journal}{Phys. Rev. Lett.}}
  \textbf{\bibinfo{volume}{101}}, \bibinfo{pages}{186805}
  (\bibinfo{year}{2008}).

\bibitem{Setty2018}
\bibinfo{author}{Phillips, P.~W.}, \bibinfo{author}{Setty, C.} \&
  \bibinfo{author}{Zhang, S.}
\newblock \bibinfo{title}{Absence of a charge diffusion pole at finite energies
  in an exactly solvable interacting flat-band model in d dimensions}.
\newblock \emph{\bibinfo{journal}{Phys. Rev. B}} \textbf{\bibinfo{volume}{97}},
  \bibinfo{pages}{195102} (\bibinfo{year}{2018}).

\bibitem{Setty2020}
\bibinfo{author}{Setty, C.}
\newblock \bibinfo{title}{Pairing instability on a luttinger surface: A
  non-fermi liquid to superconductor transition and its sachdev-ye-kitaev
  dual}.
\newblock \emph{\bibinfo{journal}{Phys. Rev. B}}
  \textbf{\bibinfo{volume}{101}}, \bibinfo{pages}{184506}
  (\bibinfo{year}{2020}).

\bibitem{Setty2021}
\bibinfo{author}{Setty, C.}
\newblock \bibinfo{title}{Superconductivity from luttinger surfaces: Emergent
  sachdev-ye-kitaev physics with infinite-body interactions}.
\newblock \emph{\bibinfo{journal}{Phys. Rev. B}}
  \textbf{\bibinfo{volume}{103}}, \bibinfo{pages}{014501}
  (\bibinfo{year}{2021}).

\bibitem{Phillips2020}
\bibinfo{author}{Phillips, P.~W.}, \bibinfo{author}{Yeo, L.} \&
  \bibinfo{author}{Huang, E.~W.}
\newblock \bibinfo{title}{Exact theory for superconductivity in a doped mott
  insulator}.
\newblock \emph{\bibinfo{journal}{Nat. Phys.}} \textbf{\bibinfo{volume}{16}},
  \bibinfo{pages}{1175} (\bibinfo{year}{2020}).

\bibitem{Yang2021}
\bibinfo{author}{Yang, K.}
\newblock \bibinfo{title}{Exactly solvable model of fermi arcs and pseudogap}.
\newblock \emph{\bibinfo{journal}{Phys. Rev. B}}
  \textbf{\bibinfo{volume}{103}}, \bibinfo{pages}{024529}
  (\bibinfo{year}{2021}).

\bibitem{Wang2021}
\bibinfo{author}{Zhu, H.-S.}, \bibinfo{author}{Li, Z.}, \bibinfo{author}{Han,
  Q.} \& \bibinfo{author}{Wang, Z.}
\newblock \bibinfo{title}{Topological s-wave superconductors driven by electron
  correlation}.
\newblock \emph{\bibinfo{journal}{Phys. Rev. B}}
  \textbf{\bibinfo{volume}{103}}, \bibinfo{pages}{024514}
  (\bibinfo{year}{2021}).

\bibitem{Setty2021-Kondo}
\bibinfo{author}{Setty, C.}
\newblock \bibinfo{title}{Dilute magnetic moments in an exactly solvable
  interacting host}.
\newblock \emph{\bibinfo{journal}{arXiv preprint arXiv:2105.15205}}
  (\bibinfo{year}{2021}).

\bibitem{Bradlyn2023}
\bibinfo{author}{Manning-Coe, D.} \& \bibinfo{author}{Bradlyn, B.}
\newblock \bibinfo{title}{Ground state stability, symmetry, and degeneracy in
  mott insulators with long range interactions}.
\newblock \emph{\bibinfo{journal}{arXiv preprint arXiv:2306.00221}}
  (\bibinfo{year}{2023}).

\bibitem{Volovik2003}
\bibinfo{author}{Volovik, G.~E.}
\newblock \emph{\bibinfo{title}{The universe in a helium droplet}}, vol.
  \bibinfo{volume}{117} (\bibinfo{publisher}{OUP Oxford},
  \bibinfo{year}{2003}).

\bibitem{Xu2015}
\bibinfo{author}{Slagle, K.}, \bibinfo{author}{You, Y.-Z.} \&
  \bibinfo{author}{Xu, C.}
\newblock \bibinfo{title}{Exotic quantum phase transitions of strongly
  interacting topological insulators}.
\newblock \emph{\bibinfo{journal}{Phys. Rev. B}} \textbf{\bibinfo{volume}{91}},
  \bibinfo{pages}{115121} (\bibinfo{year}{2015}).

\bibitem{Phillips2023}
\bibinfo{author}{Zhao, J.}, \bibinfo{author}{Mai, P.},
  \bibinfo{author}{Bradlyn, B.} \& \bibinfo{author}{Phillips, P.}
\newblock \bibinfo{title}{Failure of topological invariants in strongly
  correlated matter}.
\newblock \emph{\bibinfo{journal}{Phys. Rev. Lett.}}
  \textbf{\bibinfo{volume}{131}}, \bibinfo{pages}{106601}
  (\bibinfo{year}{2023}).

\bibitem{Goldman2023}
\bibinfo{author}{Gavensky, L.~P.}, \bibinfo{author}{Sachdev, S.} \&
  \bibinfo{author}{Goldman, N.}
\newblock \bibinfo{title}{Connecting the many-body chern number to luttinger's
  theorem through {St{\v{r}}eda's} formula}.
\newblock \emph{\bibinfo{journal}{arXiv preprint arXiv:2309.02483}}
  (\bibinfo{year}{2023}).

\bibitem{Rosch2007}
\bibinfo{author}{Rosch, A.}
\newblock \bibinfo{title}{Breakdown of luttinger's theorem in two-orbital mott
  insulators}.
\newblock \emph{\bibinfo{journal}{Eur. Phys. J. B}}
  \textbf{\bibinfo{volume}{59}}, \bibinfo{pages}{495} (\bibinfo{year}{2007}).

\end{thebibliography}

\vspace{0.8cm}

\vspace{0.3cm}
\noindent{\bf Acknowledgments}\\
We thank Roser Valenti, Cenke Xu, 
and Gabriela Oprea
for useful discussions.
Work at Rice has primarily been supported by the Air Force Office of Scientific Research under Grant No.
FA9550-21-1-0356 (conceptualization and model, C.S. and S.S.), 
by the National Science Foundation
under Grant No. DMR-2220603 (model calculation, F.X.),
by the Robert A. Welch Foundation Grant No. C-1411 (model calculation, L.C.),
and by the Vannevar Bush Faculty Fellowship ONR-VB N00014-23-1-2870 (conceptualization, Q.S.). The
majority of the computational calculations have been performed on the Shared University Grid
at Rice funded by NSF under Grant EIA-0216467, a partnership between Rice University, Sun
Microsystems, and Sigma Solutions, Inc., the Big-Data Private-Cloud Research Cyberinfrastructure
MRI-award funded by NSF under Grant No. CNS-1338099, and the Extreme Science and
Engineering Discovery Environment (XSEDE) by NSF under Grant No. DMR170109. 
S.P. acknowledges funding by the European Union (ERC, CorMeTop, project 101055088).
M.G.V. acknowledges support to the  Spanish Ministerio de Ciencia e Innovacion (grant PID2022-142008NB-I00), partial support from European Research Council (ERC) grant agreement no. 101020833 and the European Union NextGenerationEU/PRTR-C17.I1, as well as by the IKUR Strategy under the collaboration agreement between Ikerbasque Foundation and DIPC on behalf of the Department of Education of the Basque Government.
S.P. and M.G.V. acknowledge funding from the Deutsche Forschungsgemeinschaft (DFG, German Research Foundation) and the Austrian Science Fund (FWF) through the project FOR 5249 (QUAST).
J.C. acknowledges the support of
the National Science Foundation under Grant No. DMR-1942447, support from the Alfred P.
Sloan Foundation through a Sloan Research Fellowship and the support of the Flatiron Institute,
a division of the Simons Foundation. 
All authors acknowledge 
the hospitality of the Kavli Institute for Theoretical Physics, UCSB,
supported in part
by the National Science Foundation under Grant No. NSF PHY-1748958,
 during the program ``A Quantum Universe in
a Crystal: Symmetry and Topology across the Correlation Spectrum."  S.S., J.C. and Q.S. also 
acknowledge the hospitality of the Aspen Center for Physics, which is supported by the National Science Foundation under Grant No. PHY-2210452, during the workshop ``New Directions on Strange Metals in Correlated Systems."

\vspace{0.2cm}
\noindent{\bf Author contributions}\\
All authors contributed to the research of the work and the writing of the paper.

\vspace{0.2cm}
\noindent{\bf Competing 
 interests}\\
The authors declare no competing 
 interests.
 
 \vspace{0.2cm}
 \noindent{\bf Additional information}\\
Correspondence and requests for materials should be addressed to 
C.S. (csetty@rice.edu) and 
Q.S. (qmsi@rice.edu)

\clearpage

\clearpage

\clearpage

\clearpage

\setcounter{figure}{0}
\setcounter{equation}{0}
\makeatletter
\renewcommand{\thefigure}{S\@arabic\c@figure}
\renewcommand{\theequation}{S\arabic{equation}}

\noindent{\bf\Large Supplemental Information}\\
\\

\section{Lattice model} \label{app:lattice}
\begin{figure}[!t]
\centering
\includegraphics[width=1\columnwidth]{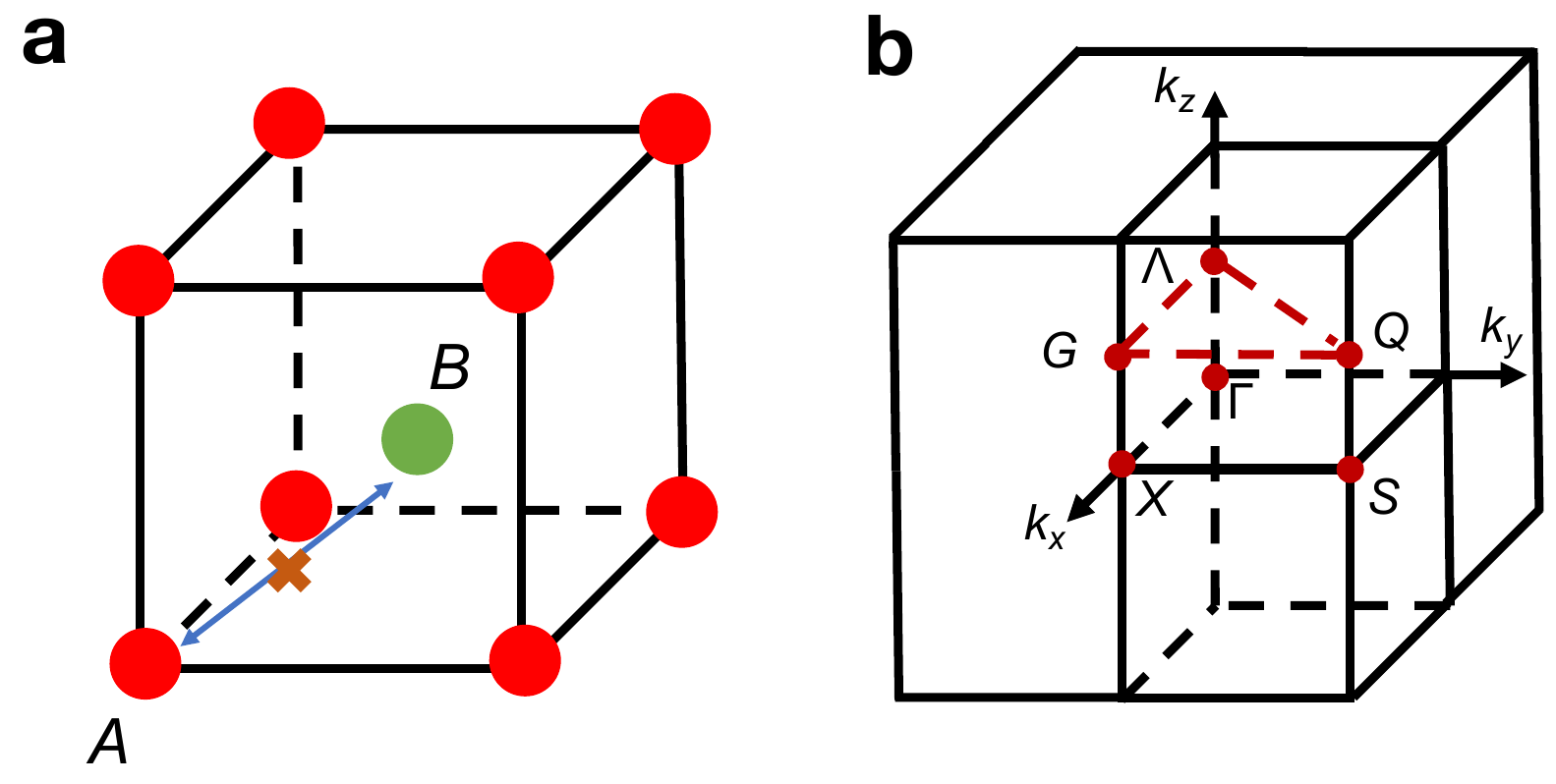}%
\caption{\textbf{Lattice structure and Brillouin zone. } \textbf{(a)} Body-centered tetragonal lattice 
used for the topological analysis in the main text. The cross mark indicates that there are no kinetic energy terms that connect the two sublattices (denoted by red and green disks). They are instead coupled through the interaction terms. \textbf{(b)} the associated Brillouin zone. The red dashed lines show the high symmetry lines used in Fig.~\ref{ZeroMonopole}(b, c) of the main text.   }
\label{fig:lattice}
\end{figure}

Here, we describe the lattice model
of our
study described
in the main text.
We consider a body-centered tetragonal lattice in Fig.~\ref{fig:lattice}.
The hoppings in the $(x,y)$-plane
are distinct from that along $\hat z$.
The 3D lattice 
has a fourfold rotational symmetry about the $\hat z$ axis, and two  sites per unit cell.
The electrons originating from each lattice site
are
spinful, and their hoppings involve spin-flipping.
We constrain the two lattice sites within a unit cell, $A$ and $B$,  to be primarily coupled  by repulsive interactions, with single-electron hoppings between these sites being negligible.

The single particle Hamiltonian, $h_0$, for each sub-lattice describes a time-reversal symmetry (TRS) broken Weyl semimetal,
\begin{align}
h_0(\bs k) = \vec n(\bs k). \vec{\sigma} - \mu \sigma_0
\end{align}
where $\sigma_\mu$ acts on the the spin degrees of freedom, and we have included a chemical potential term, $\mu$, for future convenience.
The vector~\cite{yang2011quantum, delplace2012, zyuzin2012} 
\begin{align}
\vec n(\bs k) \equiv \qty{t_x \sin k_x, t_y \sin k_y, t_0 \qty(\Delta - \sum_{i=x, y, z}\cos k_i ) },
\end{align}
with $\Delta$ controlling the location of the Weyl points.
In particular, a pair of TRS-related Weyl points occur along the $\Gamma -Z$ line in the tetragonal Brillouin zone at $k_z = \pm k^* = \pm \cos^{-1}(\Delta-2)$  for $1 < \Delta < 3$. 
Considering the standard operator for TRS, $\mathcal T \equiv i \sig_2 \mathcal K$ with $\mathcal K$ implementing complex conjugation, we see that TRS in this model is broken by a finite $n_3$.
Space inversion is implemented by $\mc P \equiv \sig_3$, and $h_0$ is invariant under $\mc P$.
We note that $h_0$ breaks the mirror symmetries on the $(k_x, k_y)$-plane; consequently, these planes can support non-trivial Chern numbers.

Since the two sublattices are not coupled by single-particle hoppings, the full Hamiltonian is given by 
\begin{align}
H_0(\bs k) = \tau_0 \otimes h_0(\bs k) \,
\end{align}
where $\tau_\mu$ acts on the sublattice degree of freedom.
Thus, our model describes a semimetallic state with doubly degenerate bands. 
The semimetal is conveniently viewed as two identical copies of Weyl semimetals, with the Weyl cones centered about the same point in the Brillouin zone carrying the same chirality.

Because all symmetry operators for $h_0$ can be straightforwardly extended to $H_0$ as $\mc X \to \tau_0 \otimes \mc X$, $H_0$ and $h_0$ share identical symmetry properties.
$H_0$ possesses more symmetries than $h_0$, however, owing to the presence of the sublattice degree of freedom.
In particular, $H_0$ is invariant under arbitrary basis transformations in the sublattice space, $e^{-i\theta \hat m \cdot \vec \tau \otimes \sig_0} H_0(\bs k) e^{i\theta \hat m \cdot \vec \tau \otimes \sig_0} = H_0(\bs k)$ with $\hat m$ being an unit 3-component vector, which leads to a sublattice-$SU(2)$ symmetry.

The presence of full ``sublattice rotation symmetry'' allows us to compute the Chern number separately for each sublattice which is equal. We then obtain the total Chern number by adding the two sublattice Chern numbers as a topological invariant for characterizing individual $k_z$ planes in the Brillouin zone. 

In the main text we have employed a second quantized form of $H_0$,
\beq
\mathscr{H}_0 = \sum_{\bs k} \Phi^{\dagger}(\bs k) \left [ \tau_0 \otimes h_0(\bs k) \right ] \Phi(\bs k) \,
\eeq
with the basis  $\Phi_{\bs k} = (
c_{A, \Uparrow}, c_{A, \Downarrow}, c_{B, \Uparrow}, c_{B, \Downarrow}
)_{\bs k}^\intercal$, where $c_{j, s, \bs k}$ destroys an electron in sublattice $j = A, B$, spin $s= \Uparrow, \Downarrow$ and momentum $\bs k$.
Without loss of generality, we have also set $t_x = t_y = t_0$, which produces a four-fold rotational symmetry that is not essential for the physics discussed here~\cite{wan2011}. 
Our formalism can be directly applied when  sublattice-$SU(2)$ symmetry is broken down to a $U(1)$ symmetry and
$\mathscr{H}_0 = \sum_{\bs k} \Phi^{\dagger}(\bs k) \left [ h_0(\bs k) \oplus h_0^*(-\bs k) \right ] \Phi(\bs k)$.
This model describes an $U(1)$-symmetric Dirac semimetal, and it is constituted by layers of the Bernevig-Hughes-Zhang model~\cite{bernevig2006} of spin-Hall insulator.
In this case,  time-reversal symmetry is preserved, and the eigenstates continue to be two-fold degenerate. 
In contrast to the previous case,  the overlapping Weyl cones carry opposite chirality, and the total Chern number for individual $k_z$ planes sums to zero.
The difference between the Chern numbers carried by the degenerate pair of bands, however, can be non-trivial, and it would replace the frequency-dependent Chern invariant defined in Eq (3) of the main text.

Further reduction of the symmetry in the sub-lattice sector, in the presence of a combined time-reversal and inversion symmetries,  would result in a non-Abelian Berry curvature.
A quantized Berry flux may be defined in such cases in the presence of rotational symmetry~\cite{tyner2023}.
Rotational  symmetry-protected Dirac semimetals~\cite{yang2014classification} exemplify this class of models.
Therefore, our overall formalism, leading to a frequency dependent quantized Berry flux, remains applicable, although the details of the topological invariant will change.

 \begin{figure}[!t]
    \centering
    \includegraphics[width=1\linewidth]{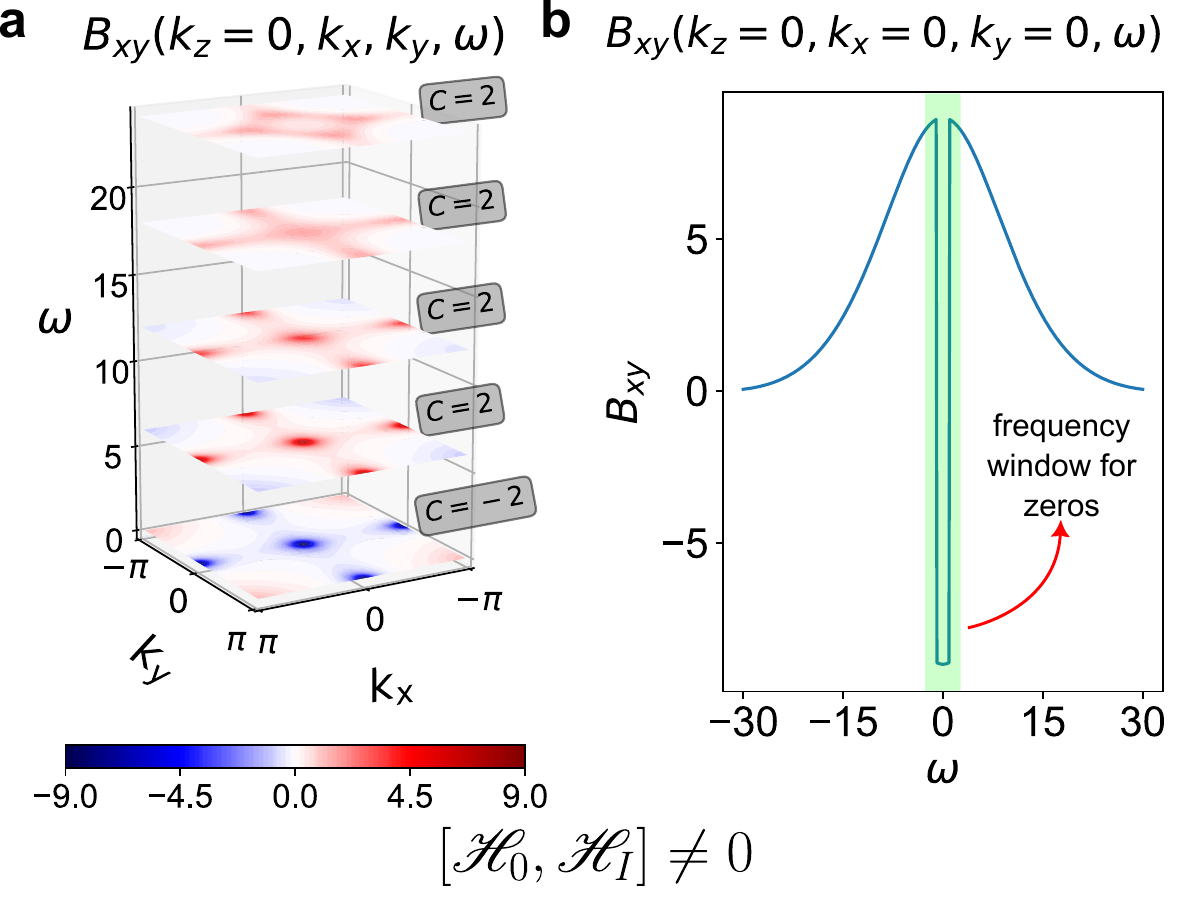}
    \caption{ \textbf{Frequency and momentum resolved Berry curvature obtained from $G_+(\bs k, \omega)^{-1}$.} Panel \textbf{(a)} shows $B_{ij}(\bs k, \omega)$ and frequency-dependent Chern number for $k_z = 0$ and various frequency slices as a function of ($k_x, k_y$). Panel \textbf{(b)} 
    displays the Berry curvature as a function of frequency for the $\Gamma$ point; the frequency window containing zeros, where the Berry curvature changes sign, is shown in the shaded green region. The non-trivial frequency dependence of the Berry curvature is reflective of the strongly interacting nature of the 
    system, in which generically the interactions do not commute with the kinetic energy (\textit{cf.} Fig.~\ref{Fig:Commute}).
    }
    \label{Fig:NoCommute}
\end{figure}

\section{Further 
details about the frequency dependent Berry curvature in the illustrative model} \label{app:FDBC}

In this section, we 
expound on the frequency dependence of the Berry curvature in the specific model we analyze. In Fig.~\ref{Fig:BerryCurvature-Main} of the main text, we showed the Berry curvature as a function of $(k_x, k_y)$ for various frequency slices with $k_z = 0, \pi$. Here, we explicitly show the frequency dependence of the Berry curvature for more clarity.  For convenience, in Fig.~\ref{Fig:NoCommute}(a)
 we again show the Berry curvature as a function of $(k_x, k_y)$ for various $\omega$ values and fixed $k_z=0$. In Fig.~\ref{Fig:NoCommute}(b) we also plot the Berry curvature as a function of frequency for a fixed point in momentum $(k_x, k_y, k_z) = (0, 0, 0)$. 
 
 Several noteworthy features are of interest. First, the plot clearly shows that the Berry curvature is strongly frequency dependent. This feature is absent in a noninteracting Hamiltonian 
and is a characteristic of an interacting 
system when generically the interactions do not commute with the kinetic energy (\textit{cf.} Fig.~\ref{Fig:Commute}). 

Second, within the frequency window between the zeros (shaded green region in Fig.~\ref{Fig:NoCommute}(b)),
the Berry curvature has the opposite sign when compared to the frequencies outside the zeros. As explained in the main text, this difference in the sign of the Berry curvature is due to the ordering of the Green's function bands which can generally differ between and outside the zero bands. Hence the eigenvectors that constitute the set $\mathscr S$ will vary accordingly. We will further touch upon this property 
later in 
the SI, Sec.~\ref{app:commuting}.  

Finally, as a note of clarification, we point out that,  for a given momentum point, the jump in the Berry curvature occurs at exactly the two frequencies where the zeros are located. In the schematic Fig.~\ref{fig:summary}, the jump is shown to occur across a band of energies (blue shaded region) as opposed to two points. This is because, in Fig.~\ref{fig:summary}, the horizontal axis is $k_z$. Hence the blue band of energies is a projection of the entire ($k_x, k_y$) plane for a given $k_z$, and for a given $k_z$, the width of the shaded region is the bandwidth of the zero bands. 

 \begin{figure*}[h]
    \centering
    \includegraphics[width=1\linewidth]{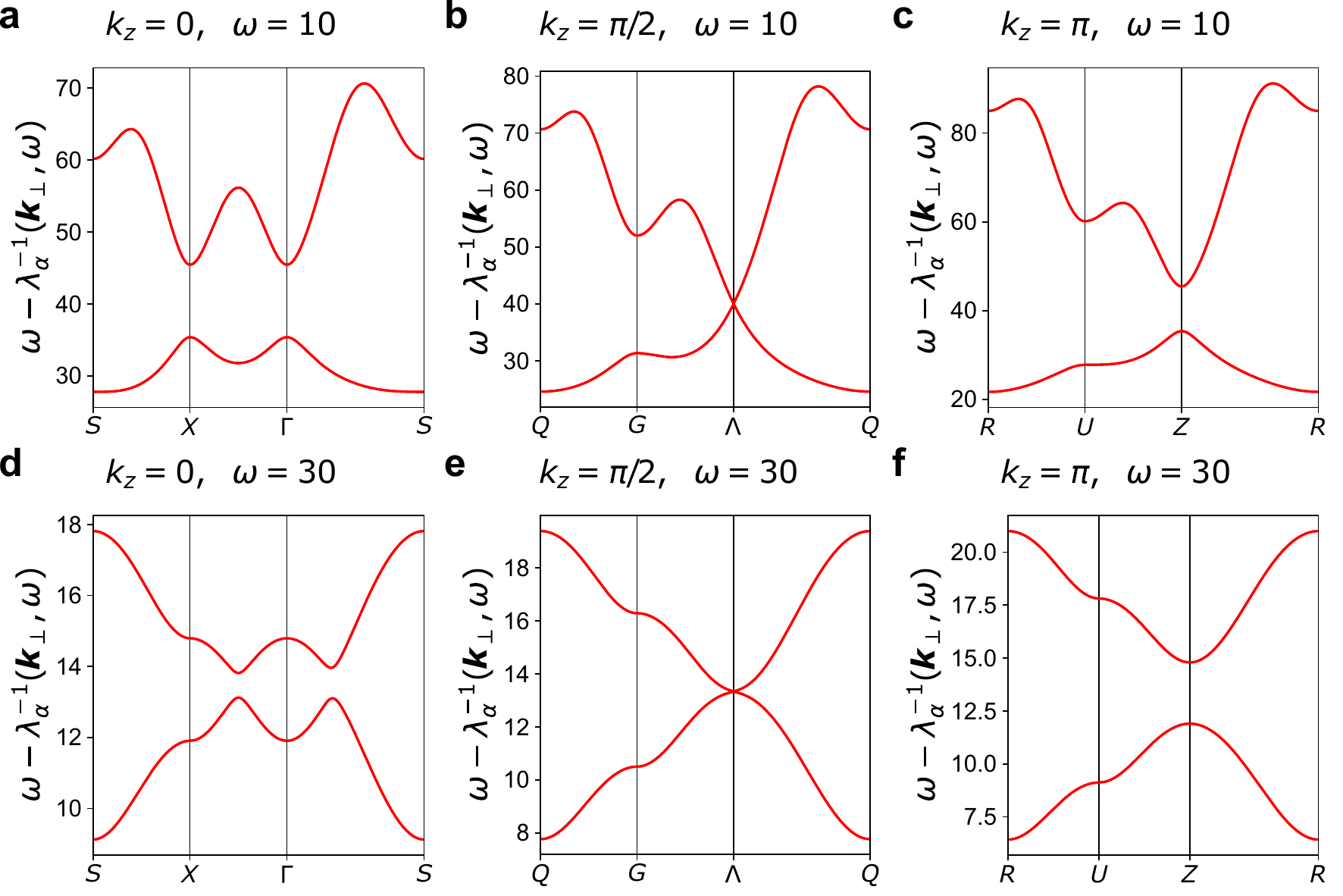}
    \caption{ \textbf{Frequency dependence of the Green's function bands.} Plots of the $\omega - G_+(\bs k, \omega)^{-1}$ `bands' (or eigenvalues of $\omega- G_+(\bs k, \omega)^{-1}$) for various values of $(\omega, k_z)$ shown in the individual panels. \textbf{(a), (b), (c)} Green's function bands for $\omega=10$ and \textbf{(d), (e), (f)} for $\omega = 30$.  The spectral crossing occurs at the critical value $k_z = k^* = \pi/2$. The bands are topological for $k<k^*$ and trivial for $k>k^*$. Note that both crossings in the center panels are linear in momentum. 
    }
    \label{Fig:BandStructureGInv}
\end{figure*}

\section{Berry curvature and frequency-dependent Chern number} \label{app:extraplots}
In this
section,
we show additional Berry curvature and frequency-dependent Chern number plots to supplement those provided in the main text.  We begin by 
displaying the eigen spectrum of $G_+(\bs k, \omega)^{-1}$ in Fig.~\ref{Fig:BandStructureGInv}. Across the columns from left to right, the various panels correspond to different $k_z$ cuts. The $k_z = 0$ ($k_z= \pi$) plane corresponds to topologically non-trivial (trivial) slice. The central column corresponds to the critical crossing point of $k_z = k^* = \pi/2$. The top (bottom) row is plotted for the frequency cut $\omega = 10$ ($\omega = 30$).  The crossing of the Green's function `bands' are generally protected by lattice symmetries~\cite{Setty2023}.

\begin{figure*}[h]
    \centering
    \includegraphics[width=1\linewidth]{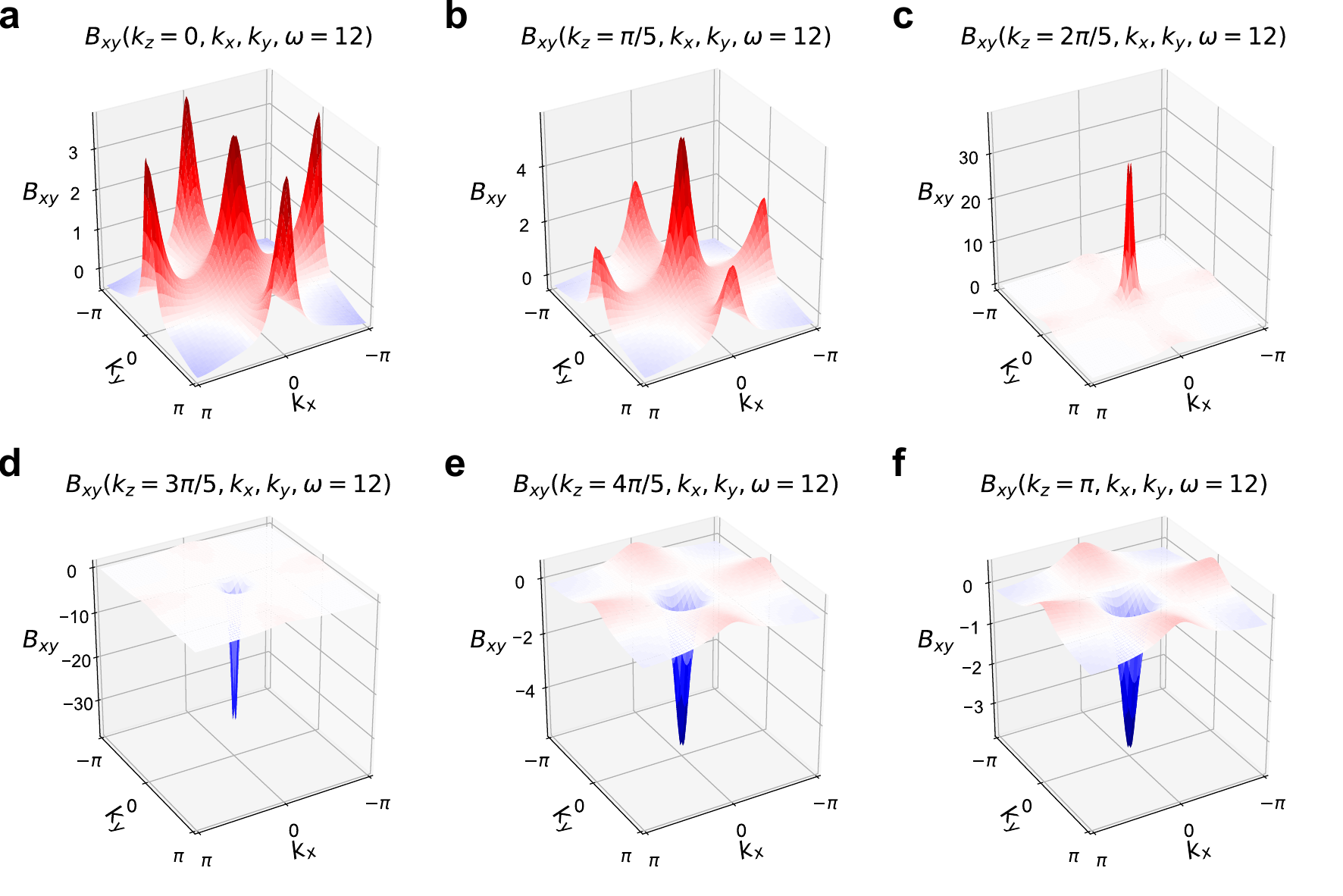}
    \caption{\textbf{Three dimensional plots of the Greeen's function Berry curvature.} Plots of $B_{ij}(\bs k, \omega)$ for a fixed frequency $\omega = 12$ are computed from Green's function eigenvectors. Different panels correspond to different cuts along $k_z$ as shown in each panel. Panels \textbf{(a-f)} correspond to $k_z = 0, \frac{\pi}{5},\frac{2\pi}{5}, \frac{3\pi}{5}, \frac{4\pi}{5}, \pi$ respectively.  Red (blue) colors in the intensity color scale denote positive (negative) Berry curvature.  Above a critical value of $k_z = k^* = \pi/2$, the interacting system is topologically trivial with $C(\omega, k_z > k^*)=0$. Below the critical value, there is a non-trivial frequency-dependent Chern number obtained from a Berry curvature that varies with frequency and $k_z$. 
    }
    \label{Fig:3DBerryCurvature}
\end{figure*}

For 
enhanced clarity, in Fig.~\ref{Fig:3DBerryCurvature} we show three dimensional plots of the Berry curvature as a function of $(k_x, k_y)$ for different slices of $k_z$. 
Here, the
frequency is fixed at $\omega = 12$. The red (blue) regions mark the positive (negative) Berry curvature. For the cases $k_z= 0, \pi/5, 2\pi/5$, the Berry curvature averages to a positive value and yields a frequency-dependent Chern number  of $+2$. However, for $k_z=  3\pi/5, 4\pi/5, \pi$, the Berry curvature averages out to produce a frequency-dependent Chern number equal to zero. 

\begin{figure*}[h]
    \centering
    \includegraphics[width=1\linewidth]{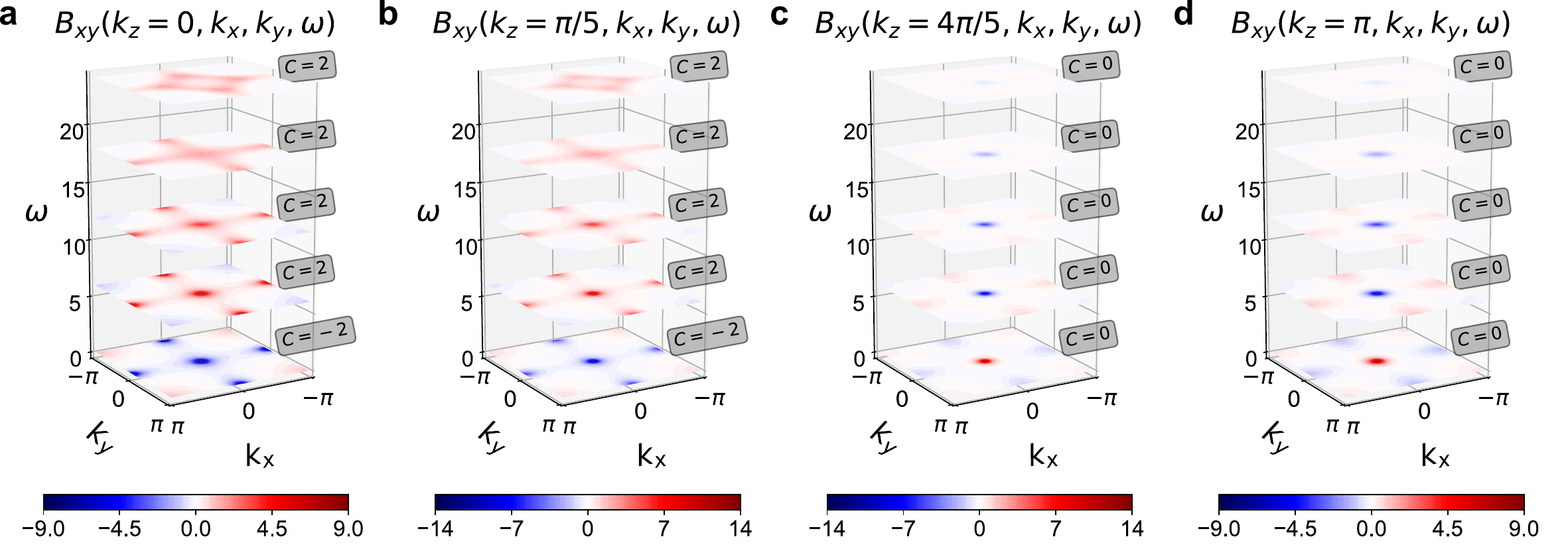}
    \caption{\textbf{Frequency and momentum resolved Berry curvature across the spectral crossing.} The Berry curvature $B_{ij}(\bs k, \omega)$,
    computed from the eigenvectors of $G_+(\bs k, \omega)^{-1}$, is plotted as a function of frequency, for different cuts along $k_z$. Panels \textbf{(a-d)} correspond to $k_z= 0, \frac{\pi}{5}, \frac{4\pi}{5}, \pi$, respectively. We have chosen the generic case where the interactions do not commute with the kinetic energy. Red (blue) colors on the intensity color scale denote positive (negative) Berry curvature. The intensity of the color denotes the magnitude of the Berry curvature. The frequency-dependent Chern number for each pair of $\omega, k_z$ is shown as well. Above a critical value of $k_z = k^* = \pi/2$, the interacting system is topologically trivial with $C(\omega, k_z> k^*)=0$. Below the critical value, there is a nontrivial frequency-dependent Chern number obtained from a Berry curvature that varies with frequency and $k_z$. Note that the frequency-dependent Chern number in between the zeros is opposite to that near the poles due to opposite phase of the kinetic Hamiltonian that appears as a pole in the self-energy. See also Fig.~\ref{Fig:3DBerryCurvature} in the SI, Sec.~\ref{app:extraplots}. 
    }
    \label{Fig:BerryCurvature}
\end{figure*}

Further in Fig.~\ref{Fig:BerryCurvature}, we plot the frequency-dependent Berry curvature and frequency-dependent Chern number, along additional $k_z$ slices, and make a comparison with the $k_z = 0, \pi$ slices shown in the main text. For intermediate values of $0<k_z<\pi$, the Berry curvature is more concentrated at the high symmetry points in the $k_x - ky$ plane (see intensity markers in Fig.~\ref{Fig:BerryCurvature}). Nonetheless, the shapes of the Berry curvature evolution for $k_z<k^* = \pi/2$ ($k_z>k^* = \pi/2$) are qualitatively similar to the $k_z = 0$ ($k_z = \pi$) scenario discussed in the main text. 
The quantization values of frequency-dependent Chern number for $k_z<k^* = \pi/2$ ($k_z>k^* = \pi/2$), however, remain the same as in the $k_z = 0$ ($k_z = \pi$) case.

\section{Homotopy invariant for $G_+$} \label{app:N-band}
Here we argue that the \emph{Hermitian} Green's function, $G_+(\bs k, \omega)$, describing sublattice-rotational symmetric interacting systems with $2N$ degrees of freedom, admits a $\mathbb{Z}^{\times(N-1)}$ classification.
Our results follow from analogous classification of $N$-band Hamiltonians with nondegenerate bands~\cite{avron1983, moore2008}.
$G_+(\bs k, \omega)$ is diagonalized by $U_G(\bs k, \omega)$ such that,
\begin{align}
G_+(\bs k, \omega) = U_G^\dagger(\bs k, \omega)  G_+^{\text{diag}}(\bs k, \omega) U_G(\bs k, \omega),
\end{align}
where $G_+^{\text{diag}}(\bs k, \omega)$ is a diagonal matrix.
Due to the $SU(2)$ sublattice rotational symmetry, $G_+ = \tau_0 \otimes \mc G_+$, and $U_G$ acquires a block-diagonal form with each block being the same $N\times N$ unitary matrix, $u_G \in U(N)$, which diagonalizes $\mc G_+$.
Since the eigenvalues of $\mc G_+$ are nondegenerate at generic points of the $(\bs k, \omega)$ space, $u_G$ is defined up to $U(1)^{\times N} \coloneqq \underbrace{U(1) \times \ldots \times U(1)}_{\text{N copies}}$ gauge transformations, such that replacing $u_G \to u_G W$, with $W \in U(1)^{\times N}$, leaves $G_+^{\text{diag}}$ invariant. 
Thus, the existence of a topological obstruction in the eigenvectors of $\mc G_+$ is diagnosed by the homotopy group of the coset space $U(N)/U(1)^{\times N} \equiv SU(N)/U(1)^{\times (N-1)}$.
It is known~\cite{avron1983} that
\begin{align}
\pi_2\qty(\frac{SU(N)}{U(1)^{\times (N-1)}}) = \mathbb Z^{\times (N-1)},
\end{align}
which implies the existence of two-dimensional subspaces of the $(\bs k, \omega)$-space in which the eigenstates of $\mc G_+$ can  support a quantized Berry flux.
Since $G_+$ is constituted by two identical copies of $\mc G_+$, the net Chern number for each eigenstate of $G_+$ is twice that of $\mc G_+$.
In the main  
text, we demonstrated the existence of such a quantized Berry flux on $(k_x, k_y)$-planes of the Brillouin zone for a specific model.
We note that this classification scheme avoids the notion of ``filled'' vs. ``empty'' eigenvectors of Green's functions.

When the sublattice-$SU(2)$ is broken down to a $U(1)$ symmetry, the Green's function takes the form $G_+ = \tau_0 \otimes \mc G_+^{(0)} + \tau_3 \otimes \mc G_+^{(3)}$, where $\mc G_+^{(\mu)}$ are $N\times N$ Hermitian matrices.
Therefore, the topological classification of the eigenvectors of $\mc G_+^{(0)} \pm \mc G_+^{(3)}$ continues to guide the topological classification of the eigenvectors of $G_+$.
Here, the degenerate eigenstates carry distinct sublattice quantum numbers; consequently, their respective Chern numbers can be resolved straightforwardly.
However, the net Chern number of the degenerate pair of bands may no longer be a suitable topological invariant and would be replaced by other indices such as
spin Chern number or the $\mathbb Z_2$ index.

Finally, we also note that, in the absence of the twofold degenerate eigenvalues, $G_+$ would become analogous to $\mc G_+$, and the above topological classification can be directly applied to the eigenstates of $G_+$.
This treatment would be appropriate for a strongly correlated Weyl semimetal.

\begin{figure}[!t]
\centering
\includegraphics[width=0.95\columnwidth]{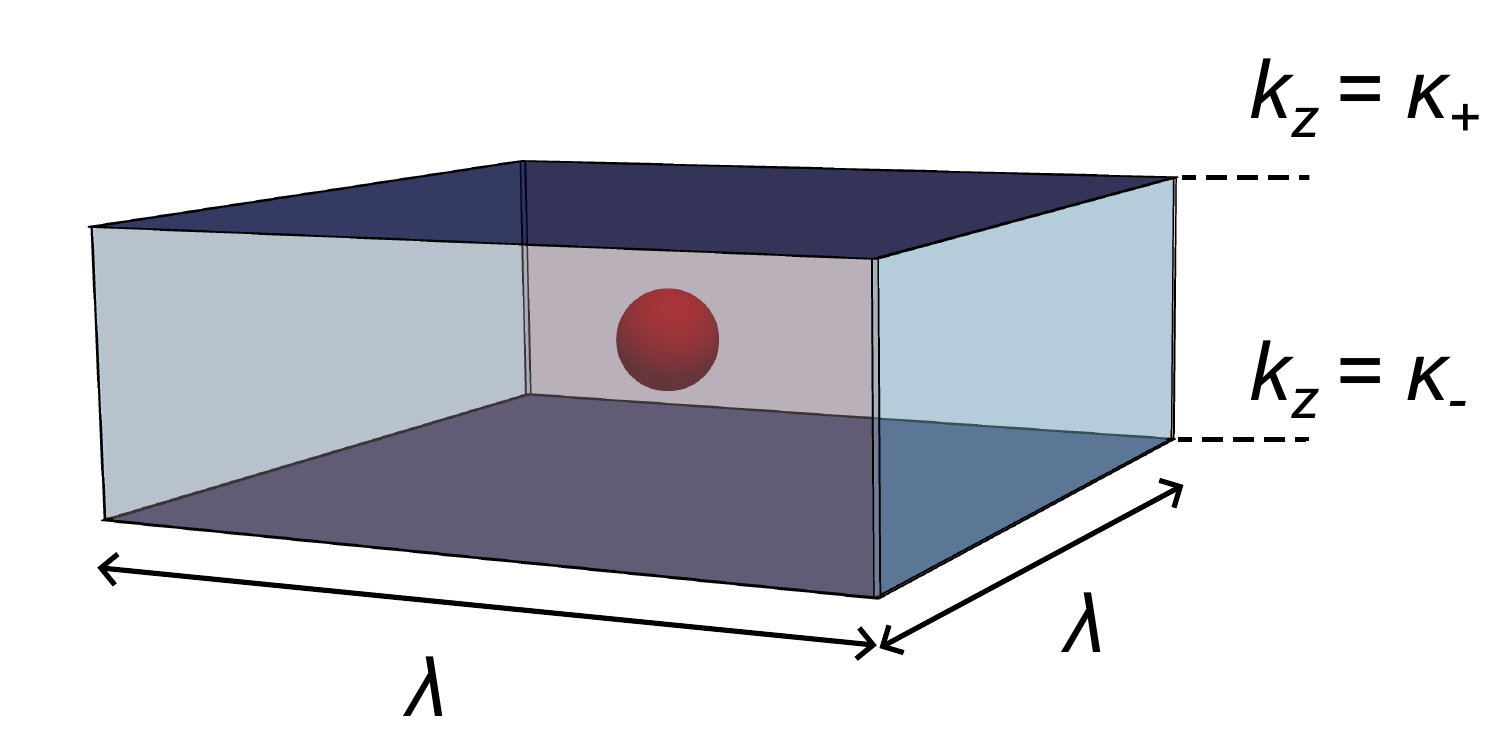}
\caption{\textbf{Quantization of
the Green's function Berry flux.} A Berry monopole (red sphere) enclosed by a pillbox shaped Gaussian surface of linear dimension $\lambda$ on the $(k_x, k_y)$ plane.
The $k_z = \kappa_\pm$ planes form the ``top'' and ``bottom'' surfaces of the pillbox.
In systems where $k_z$ planes support nontrivial Berry curvatures, the $\kappa_\pm$ surfaces of the pillbox will carry unquantized Chern numbers.
Since the pillbox Gaussian surface encloses a Berry monopole, the net Berry flux that passes through it is quantized in units of $2\pi$.
In the limit $\lambda \to \infty$ (or $2\pi$ with periodic boundary conditions), this Berry flux is entirely carried by the $\kappa_\pm$ planes, such that it has to equal the net Chern number carried by these two planes.
}
\label{fig:pillbox}
\end{figure}

\section{Band-crossing and monopoles} \label{app:pillbox}
When a pair of Chern-number carrying bands cross, their respective Chern numbers change, but the total Chern number of these bands must be conserved.  
Thus, the change in the Chern number of one of the crossing bands (say, the ``lower'' band) must be opposite to that of the other.
The magnitude of the jump in the Chern number of the lower band reveals the magnitude of the Berry monopole-charge realized at the band crossing point~\cite{Armitage2017}.

In order to connect the jump in the Chern number to the existence of a Berry monopole, it is convenient to consider a continuum version of the model of interest, and enclose the band crossing point by a Gaussian surface shaped like a `pillbox', as depicted in Fig.~\ref{fig:pillbox}.
The net Berry flux that pierces the pillbox is quantized, thanks to the Berry monopole enclosed by it.  
The surfaces of the pillbox with normals parallel/antiparallel to the $\hat k_z$-axis, being subregions of corresponding $k_z$ planes, generically support finite but unquantized Berry flux.
For a pillbox of linear dimension $\lambda$ in the $(k_x, k_y)$-plane, the net flux passing through the two planes along $\hat k_z$ (say) is given by
\begin{align}
\delta \mathscr F(\lambda) = \mathscr F_+(\lambda) - \mathscr F_-(\lambda),
\end{align}
where $\mathscr F_\pm$ are fluxes through the $k_z = \kappa_\pm$ surfaces of the pillbox (see Fig.~\ref{fig:pillbox}).
Thus, in the limit $\lambda\to \infty$ (or equivalently 2 $\pi$ when there are periodic boundary conditions),  $\delta \mathscr F(\lambda)$ is the difference between the Chern numbers carried by the two  planes at $k_z = \kappa_\pm$.

Although the total Berry flux due to the monopole passing through the pillbox remains fixed, independent of the magnitude of $\lambda$, as $\lambda$ increases the surfaces of the pillbox with normals perpendicular to the $\hat k_z$ axis progressively  carry lower fraction of the total flux.
Consequently, in the $\lambda\to \infty$ limit, the entire Berry flux due to the monopole is obtained from the $\pm$ surface of the pillbox.
Therefore, it must be equal to $\dl \mathscr F(\lambda \to \infty)$, which is also the difference between the Chern numbers of the two $k_z$ planes on either side of the band crossing point.

\section{Green's function in the commuting limit} \label{app:commuting}

In this section, we study scenarios when $\mathscr H_0$ and $\mathscr H_I$ commute with each other. We will demonstrate that the quantization condition and frequency dependence of the 
Green's function
Chern number are already captured in this limit. However, the Berry curvature acquires additional frequency dependent structure when $\mathscr H_0$ and $\mathscr H_I$ do not commute with each other. We study two specific cases that are of interest. In the first 
case, we take a special limit of $h_0(\bs k)$ that appears in the main text such that it commutes with $\mathscr H_I$.
In the second
case, we modify $\mathscr H_I$ such that it commutes with $h_0(\bs k)$.

\noindent 
\textit{Commuting case 1:}  We consider the noninteracting 
Hamiltonian $\mathscr{H}_0$ from the main text given by
$\mathscr{H}_0 = \sum_{\bs k} \Phi^{\dagger}(\bs k) \left [ \tau_0 \otimes h_0(\bs k) \right ] \Phi(\bs k)$. For analytical tractability,  we choose the limit where $\vec n(\bs k) \simeq \{0, 0, \Delta - \sum_{i=x, y, z}\cos k_i \}$, which becomes exact on the $k_z$ axis. 
In this limit,
$\mathscr H_0$ and $\mathscr H_I$ commute.
Consequently,
the noninteracting Bloch functions diagonalize the \textit{total} Hamiltonian and Green's function. These Bloch functions can then be used to compute the Berry curvature of the fully interacting system exactly. It thus serves as
reference setting for
the complete solution as is done numerically in the main text. 

To obtain these eigenvectors, we diagonalize the Green's function in the orbital basis (where $\omega$ is real frequency)
\begin{align}
G(\bs k, \omega) = \frac{1}{\tilde G^{-1}(\bs k, \omega) - (U/2)^2 \tilde G(\bs k,\omega),
}
\label{Eq:G1z}
\end{align}
by setting $|\phi_{\alpha}(\bs k, \omega)\rangle$ as the noninteracting Bloch functions. Here $\tilde G^{-1}(\bs k, \omega) \equiv \omega  \mathbbm{1} - \qty[\tau_0 \otimes h_0(\bs k) + U' \mathbbm{1} ]$ and $(U, U') = (U_c + U_s, U_c-U_s)$. We show below that the same basis diagonalizes the total Hamiltonian for large $U, U'$.

To demonstrate that the basis that diagonalizes the Green's function in Eq.~\ref{Eq:G1z} also diagonalizes the total Hamiltonian $\mathscr H$ in the strong coupling (Mott) limit, we begin by choosing the basis that diagonalizes the noninteracting bands denoted by $\Psi_{\bs k} = (
\psi_{1, \uparrow}, \psi_{1, \downarrow}, \psi_{2, \uparrow}, \psi_{2, \downarrow}
)_{\bs k}^\intercal$. Here, $\{\psi_{j, \uparrow}, \psi_{j, \downarrow}\}$ represents the $j$-th pair of degenerate bands. The noninteracting Hamiltonian is then diagonalized to  $\qty[- \mu \mathbbm{1} + |\vec n(\bs k)| \tau_3\otimes \sigma_0]$. 
Henceforth, we treat $\sigma = \uparrow, \downarrow$ and $i, j =1,2$ as a pseudo-spin and band indices respectively. 
The total Hamiltonian can then be recast into the form $H = H_0 + H_I$ where
\beq
H_0 &=& \sum_{\bs k i \sigma} \xi_{i \sigma}(\bs k) \psi^{\dagger}_{i\sigma}(\bs k)\psi_{i\sigma}(\bs k)  \\
H_I &=& U \sum_{\bs k i} n_{\bs k i \uparrow} n_{\bs k i \downarrow}  + U' \sum_{\substack{\bs k \sigma \sigma'\\ i\neq j}} n_{\bs k i \sigma} n_{\bs k j \sigma'} 
\label{Eq:BandBasisHamiltonian}
\eeq
and $(U, U') = (U_c + U_s, U_c-U_s)$ correspond to  intra-band and inter-band interactions respectively. 
$n_{\bs k i \sigma}$ denotes the density operators of $\psi_{\bs k i \sigma}$ and the band dispersions (renormalized) have the same spin dispersions $\xi_{i\uparrow}(\bs k) = 
\xi_{i \downarrow} (\bs k)$. The model written above in the band basis is the multiband version of the Hatsugai-Kohmoto (HK) model~\cite{HK1992,Nagaosa2016, Setty2018, Setty2020, Setty2021, Phillips2020, Yang2021, Wang2021, Setty2021-Kondo,Fabrizio2022, Bradlyn2023}. 
The total Green's function can be evaluated exactly by choosing $|\phi_{\alpha}(\bs k, z)\rangle$ as the Bloch basis where $z$ is the complex frequency.  To show this~\cite{Setty2023}, we note that when $\xi_1(\bs k), \xi_2(\bs k)$ are filled with $U, U'>0$, the Green's function simplifies in the zero temperature limit. 
For each $\bs k$, in the strong interaction limit when $U>2|\xi_1|, 2 |\xi_2|$ and $U + 2 U'> |\xi_1| + 2 |\xi_2|, 2|\xi_1| + |\xi_2|$,  while $U'< |\xi_1| + |\xi_2|$, the partition function close to zero temperature can be shown to take the form $Z_{\bs k} = \lim_{\beta \rightarrow \infty} 4 e^{-\beta (\xi_1 + \xi_2 + U')}$~\cite{Setty2023}. We can then obtain a pair of two fold degenerate  Green's function `bands' as  (where $z$ is a complex frequency)
\beq
w_{\pm}(\bs k, z) =  \frac{1}{2} \sum_{s=\pm} \frac{1}{z  - U' \pm |\vec n(\bs k)| + s U/2} \,.
\eeq
Here we have distinguished the Green's function `bands' ($w_{\pm}(\bs k, z)$) with the eigenvalues of $G_+(\bs k, \omega)$ ($\lambda_{\alpha}$). In the commuting case studied here, $\lambda_{\alpha}(\bs k, \omega) = \frac{1}{2}(w_{\alpha}(\bs{k}, \omega +i\eta) + w_{\alpha}(\bs{k}, \omega -i\eta))$, with $\alpha = +/-$. Therefore, we can write the self-energy in the diagonalized basis as 
\beq
\Sigma(\bs k, z) = \frac{U^2}{4} \text{diag} \{\frac{1}{z - U' + n(\bs k)}, \frac{1}{z - U' - n(\bs k)} \} \,.
\eeq
The eigenvalues $w_{\pm}(\bs k, z)$ above and the Bloch eigenstates are precisely those that are obtained by diagonalizing the Green's function in the orbital basis (after absorbing $\mu$ into $U'$ ) in Eq.~\ref{Eq:G1z}. \par 
We can now evaluate the spectral function which contains both poles and zeros of the Green's function. The poles (zeros) occur due to the existence of the first (second) term in the denominator in Eq.~\ref{Eq:G1z}. 
Since the Bloch functions are independent of frequency in this special limit and can diagonalize the full interacting Green's functions, $G(\bs k, \omega)$ and $G_+(\bs k, \omega)$,
they can be used to compute Berry curvature and frequency-dependent Chern number (according to Eqs.~\ref{Eq:BerryCurvature}, ~\ref{Eq:SpinWinding} of the main text) for both the lower/upper Hubbard bands and the zeros \textit{alike}. This is done by choosing the appropriate frequency window of interest in Fig.~\ref{ZeroMonopole} (a) of the main text. 
Thus we can conclude from this simple scenario that the frequency-dependent Chern number of zero crossings is quantized with a magnitude equal to those obtained from noninteracting $|\phi_{\alpha}(\bs k, \omega)\rangle$. It implies that the crossings of zeros behave as sources and sinks of Berry curvature similar to poles despite lacking Landau quasiparticles, as shown in Fig.~\ref{ZeroMonopole} (a).

From the Green's function in the orbital basis, it is further possible to see why the frequency-dependent Chern number near the zeros can be different from the frequency-dependent Chern number near the poles. As emphasized in the main text, the Berry curvature $B_{ij}(\bs k, \omega)$ and frequency-dependent Chern number are computed using the eigenvectors $|\phi_{\alpha}(\bs k, \omega) \rangle$. Here $\alpha \in \mathscr S$ where the set $\mathscr S$ contains the set of bands whose topological properties are of interest. In the model exemplified in the main text, we choose $\mathscr S$ to contain the lowest pair of eigenvectors of the Green's function. This set is, however, frequency dependent. As is evident from Eq.~\ref{Eq:G1z}, there is a relative negative sign in the denominator. Due to this, the ordering of the eigenvalues near the poles of the Green's function is opposite to the ordering of that near the zeros. Hence, this hierarchical arrangement of the Green's function 'bands' near the poles (first term in the denominator of Eq.~\ref{Eq:G1z} dominates) can differ from that near the zeros (second term in the denominator of Eq.~\ref{Eq:G1z} dominates). Thus the choice of the lowest two eigenvalues and eigenvectors in $\mathscr S$ is frequency dependent and so is the frequency-dependent Chern number.  
All these illustrate the point that
the frequency dependence of the Berry curvature and Chern number is a correlation driven effect.

 \begin{figure}[h]
    \centering
    \includegraphics[width=1\linewidth]{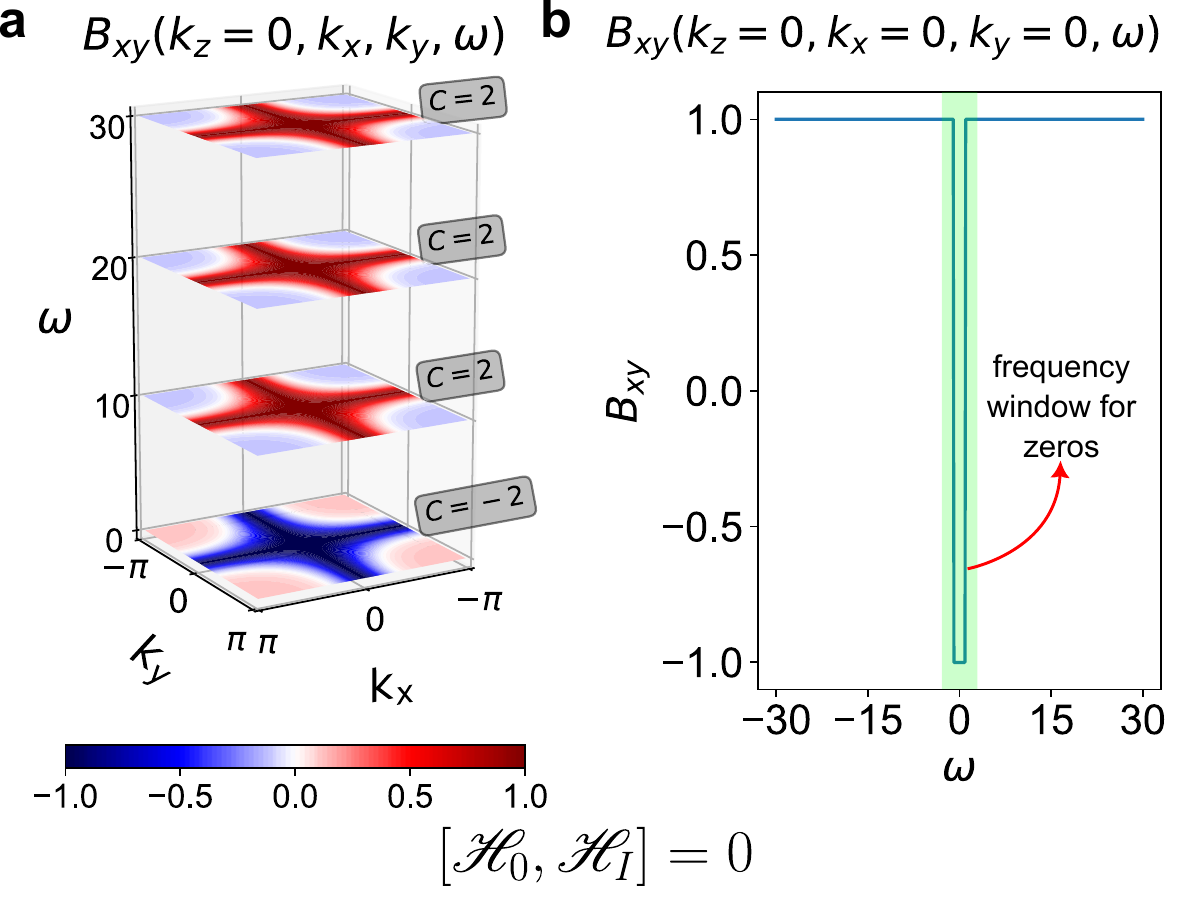}
    \caption{ \textbf{Frequency and momentum dependent Berry curvature for commuting interactions.} Plots of the Berry curvature obtained from $G_+(\bs k, \omega)^{-1}$ as a function of frequency and momentum cuts in the Brillouin zone when the interactions commute with the 
    noninteracting Hamiltonian. Panel \textbf{(a)} shows $B_{ij}(\bs k, \omega)$ and frequency-dependent Chern number for $k_z = 0$ and various frequency slices as a function of ($k_x, k_y$). Panel \textbf{(b)} shows the Berry curvature as a function of frequency for the $\Gamma$ point; the frequency window containing zeros where the Berry curvature changes sign is shown in the shaded green region.  The only frequency dependence of $B_{ij}(\bs k, \omega)$  arises between the poles and zero bands; otherwise it is featureless (\textit{cf.} Fig.~\ref{Fig:NoCommute}). Note that the frequency-dependent Chern number and its quantization for both poles and zeros are captured already  when $[\mathscr H_0, \mathscr H_I']= 0$.
    }
    \label{Fig:Commute}
\end{figure}

\textit{Commuting case 2:} To complement results of the previous 
subsection, we also briefly describe the case when we choose a four fermion term $\mathscr H_I'$ such that it commutes with the original $\mathscr{H}_0 = \sum_{\bs k} \Phi^{\dagger}(\bs k) \left [ \tau_0 \otimes h_0(\bs k) \right ] \Phi(\bs k)$. This 
takes the form 
\beq
\mathscr H_I' 
= \frac{\alpha}{2} \sum_{\bs k}  \Phi_{\bs k}^\dagger  \Phi_{\bs k} + \frac{U_c}{2} \sum_{\bs k} \left( \Phi_{\bs k}^\dagger  \Phi_{\bs k}\right)^2 
+ \frac{U_s'}{2} \sum_{\bs k} \qty( \Phi_{\bs k}^\dagger \tau_3 \otimes \sigma_0 \Phi_{\bs k})^2.
 \eeq
Unlike the $U_s$ term described in the main text,
$\mathscr H_I'$ 
commutes with $\mathscr H_0$. 
We can then use the same procedure
as described in the previous 
subsection 
to obtain the Green's function by diagonalizing the entire $\mathscr H' \equiv \mathscr H_0 + \mathscr H_I'$. Using the resultant eigenvectors $|\phi_{\alpha}(\bs k, \omega) \rangle$, it is straightforward to obtain the Berry curvature and Chern number from Eqs.~\ref{Eq:BerryCurvature}~\ref{Eq:SpinWinding} of the main text. 

Fig.~\ref{Fig:Commute} shows the momentum and frequency dependence of the computed Berry curvature and Chern number 
on the $k_z =0$ plane. The only frequency dependence of the Berry curvature and Chern number appears between the pole (red intensity) and zero bands (blue intensity). Otherwise, the Berry curvatures evolves in a featureless manner as a function of frequency (Fig.~\ref{Fig:Commute} (b)). On the other hand, when the four fermion term does not commute with the kinetic energy (panels (a,b) in Fig.~\ref{Fig:NoCommute} corresponding to $k_z= 0$), the Berry curvature evolves continuously and non-trivially with frequency. Note however that the quantization of frequency-dependent Chern number is already captured in the commuting limit.

 \begin{figure*}[h]
    \centering
    \includegraphics[width=1\linewidth]{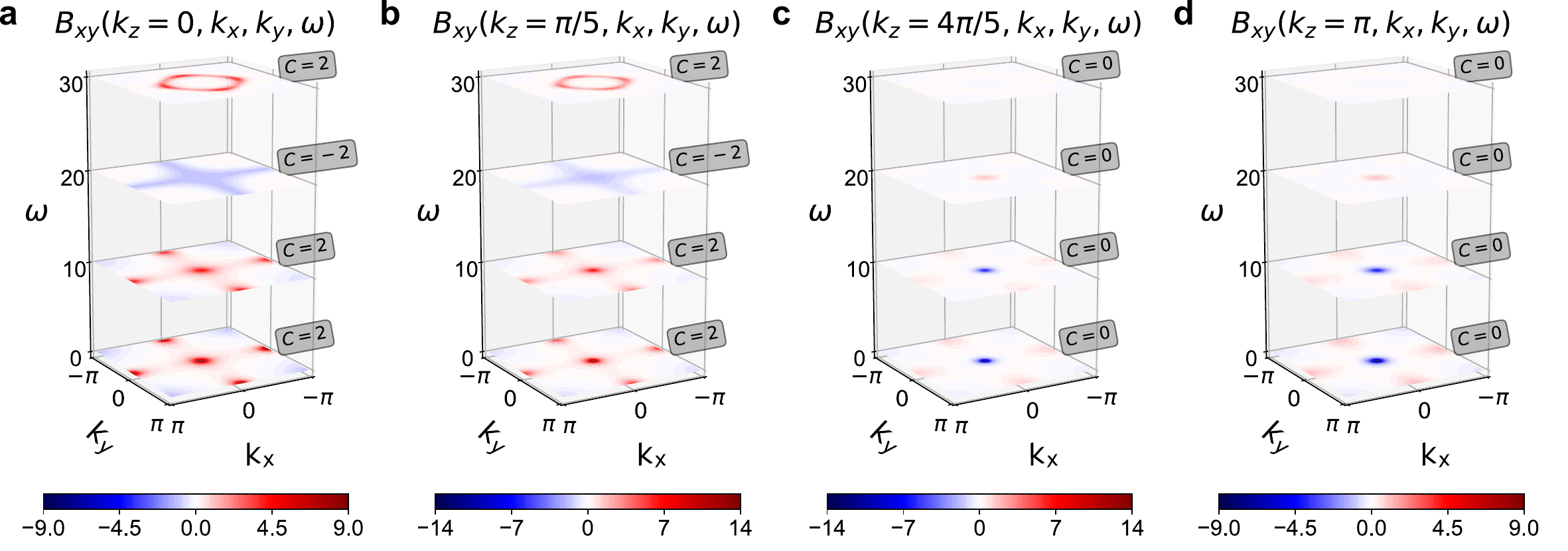}
    \caption{\textbf{Frequency and momentum resolved Berry curvature across spectral crossing from $G_+(\bs k, \omega)$. }Plots of the Berry curvature $B_{ij}(\bs k, \omega)$ as a function of frequency for different cuts along $k_z$ computed from the eigenvectors of $G_+(\bs k, \omega)$. Like in the main text, the red and blue colors in the intensity color scale denote positive and negative Berry curvature respectively.  The frequency-dependent Chern number for each pair of $(\omega, k_z)$ is also indicated. Panels \textbf{(a-d)} correspond to $k_z= 0, \frac{\pi}{5}, \frac{4\pi}{5}, \pi$, respectively.  Above a critical value of $k_z = k^* = \pi/2$, the interacting system is topologically trivial with $C(\omega, k_z> k^*)=0$. Below the critical value, there is a nontrivial frequency-dependent Chern number obtained from a Berry curvature that varies with frequency and $k_z$. }
    
    \label{Fig:BerryCurvatureG}
\end{figure*}

 \begin{figure*}[!t]
    \centering
    \includegraphics[width=1\linewidth]{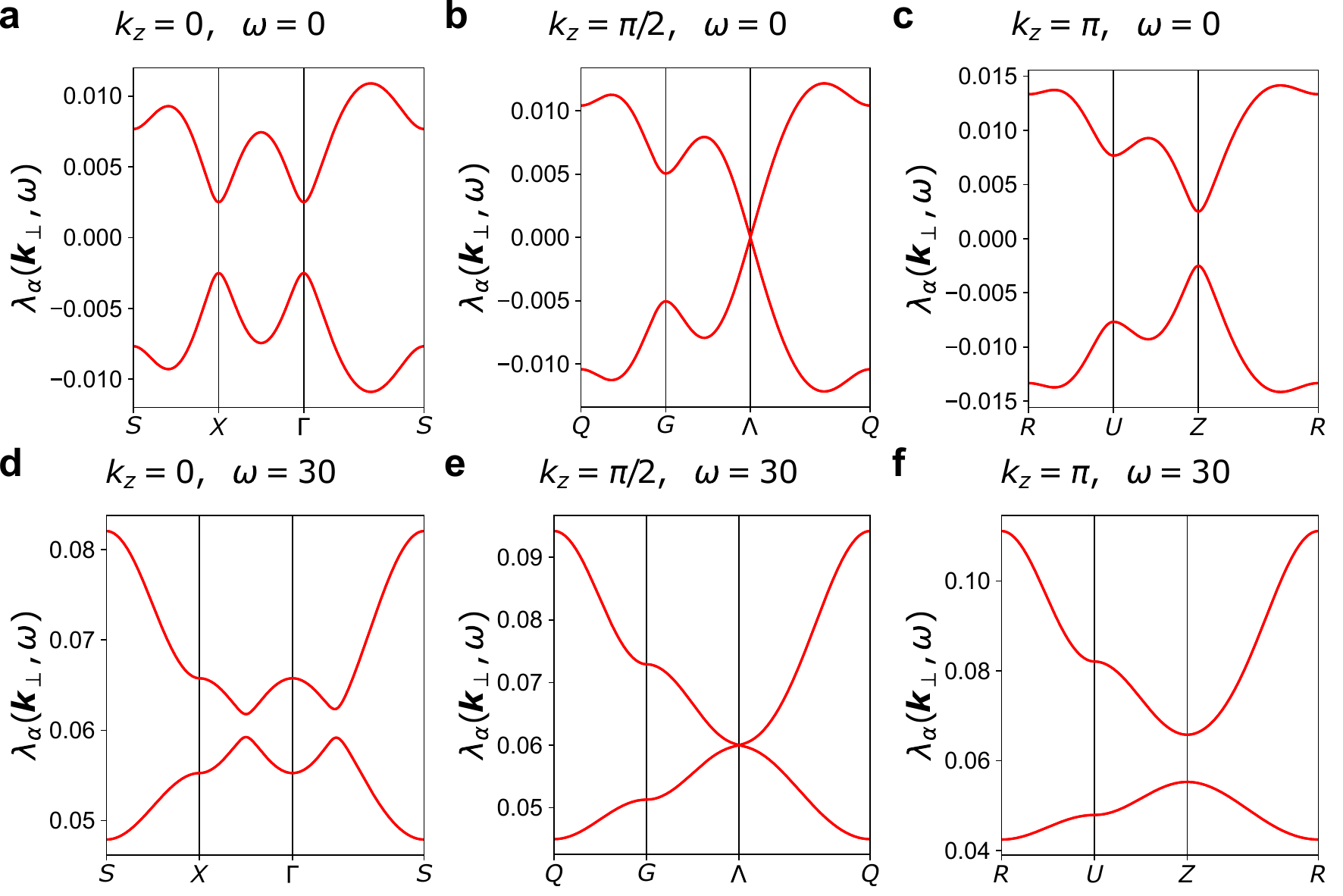}
    \caption{\textbf{Frequency dependence of Green's function bands from $G_+(\bs k, \omega)$.} Plots of the $\omega- G_+(\bs k, \omega)$ 'bands' (or eigenvalues of $\omega-G_+(\bs k, \omega)$) for various values of $(\omega, k_z)$ shown in the individual panels. \textbf{(a), (b), (c)} Green's function bands for $\omega=0$ and \textbf{(d), (e), (f)} for $\omega = 30$. The spectral crossing occurs at 
    $k_z = k^* = \pi/2$. The bands are topological for $k<k^*$ and trivial for $k>k^*$. Note that both crossings in the center panels are linear in momentum. 
    }
    \label{Fig:BandStructureG}
\end{figure*}

\section{Topological properties of $G_+(\textbf k, \omega)$ } \label{app:Gplus}
In the main text, we focused on the topological properties of $G_+(\textbf k, \omega)^{-1}$ where the poles are well behaved but the zeros diverge. In this SI, we instead study the alternative matrix $G_+(\textbf k, \omega)$ where the zeros are well behaved but the poles diverge. We show that the eigenvectors of $G_+(\textbf k, \omega)$ can also be used to characterize topology in interacting settings. This is a complementary picture that can additionally lend the interpretation of zero crossings as sources and sinks of Berry curvature. 

We follow the same procedure of computing the eigenvectors of $G_+(\textbf k, \omega)$ and using them to determine the frequency dependent Berry curvature and the Chern number. In Fig.~\ref{Fig:BerryCurvatureG}, we plot the frequency dependent Berry curvature and Chern number obtained from the eigenvectors of $G_+(\textbf k, \omega)$ for various $k_z$ cuts. Fig.~\ref{Fig:BandStructureG} shows the associated `band structure' of $G_+(\textbf k, \omega)$ for two frequencies ($\omega = 0; \omega=30$) and different $k_z$ cuts of $k_z =0$ (topological), $k_z = k^*$ (critical) and $k_z = \pi$ (trivial).  The Berry curvature obtained from the eigenvectors of $G_+(\textbf k, \omega)^{-1}$ can be contrasted with that obtained from the eigenvectors of $G_+(\textbf k, \omega)$. In the former case,  the Berry curvature for frequencies in-between the zero bands have the opposite sign compared to those outside. However, this result is flipped for the latter, i.e., the Berry curvature for frequencies in-between the pole bands (that occur for $\omega \sim 20$) have the opposite sign instead. This distinction occurs because the ordering of the eigenvalues of $G_+(\textbf k, \omega)$ and $G_+(\textbf k, \omega)^{-1}$ (and hence the composition of the set $\mathscr S$ that constitutes $| \phi_{\alpha} (\bs k, \omega) \rangle$) is different.   Nevertheless, there is a non-trivial frequency dependent Berry curvature, and the Chern number is quantized just like what was found from analysis of  $G_+(\textbf k, \omega)^{-1}$.

\begin{figure}[h]
\centering
\includegraphics[width=0.9\columnwidth]{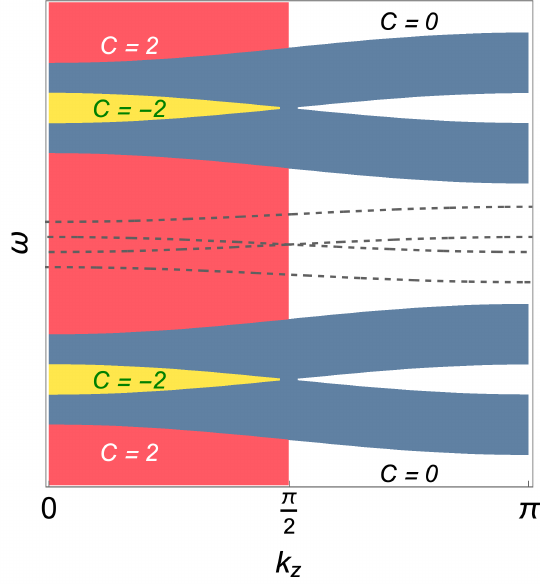}
\caption{ \textbf{Frequency dependent Chern numbers obtained from $G_+(\bs{k}, \omega)$. } Schematic representation of the frequency dependent Chern number distribution obtained from the eigenstates of $G_+(\bs k, \omega)$. The shaded region in blue now indicates bands of poles where $G_+(\bs k, \omega)$ is ill-defined. 
This is to be compared 
with Fig.~\ref{fig:summary} appearing in the main text. The noninteracting bands form a Weyl point at $\mathbf k = \mathbf k^* \equiv  (0,0,\pi/2)$ and the zero bands are enclosed by the dashed lines. }
\label{Fig:schematic_GP}
\end{figure}

In  Fig.~\ref{Fig:schematic_GP}, we show a numerically obtained schematic of the frequency-dependent Chern numbers for different continuous values of $(\omega, k_z)$ obtained from the eigenvectors of $G_+(\bs k, \omega)$.  This figure is an analog of Fig.~\ref{fig:summary} obtained for $G_+(\bs k, \omega)^{-1}$ in the main text; however, the shaded region in blue now indicates bands of poles where $G_+(\bs k, \omega)$ is ill-defined (\textit{cf.} Fig.~\ref{fig:summary} where the shaded region indicates bands of zeros where $G_+(\bs k, \omega)^{-1}$ is ill-defined). The dashed lines in Fig.~\ref{Fig:schematic_GP} enclose the projection of zero bands for a given $k_z$ plane. Just like for $G_+(\bs k, \omega)$ case, above (below) the critical value of $k_z = \pi/2$, the interacting system is topologically trivial (non-trivial) for all frequencies.  We can further conclude from the jump of frequency-dependent Chern number across the critical $k_z$ that both zeros and poles act as sources and sinks of Berry curvature. 
For a given $k_z$, we similarly find a non-trivial frequency dependence of the quantization value. If the frequency-dependent Chern number is $+C$ near zero bands (red center enclosed by dashed lines in Fig.~\ref{Fig:schematic_GP}), its value switches to $-C$ in vicinity of the poles (sandwiched yellow region in Fig.~\ref{Fig:schematic_GP}). This switch is ensured by the
 arrangement of eigenvalues of $G_+(\bs k, \omega)$ -- the eigenvalue ordering near the zeros is different from the poles. Therefore, correlation effects lend a non-trivial structure to the frequency dependent Berry curvature and Chern number.

 \section{Physical Properties}
 \label{app:PhysicalProperties}
 The question of how topological indices, well-established in the realm of 
noninteracting
 electronic systems, can be generalized in the presence of strong correlation effects is 
 well studied~\cite{Volovik2003}. In the extreme correlation limit when the self-energy diverges, several authors~\cite{Dzyaloshinskii2003, AGD, sym_green, Gurarie2011-2, Xu2015, Yunoki2017, wagner2023mott,  Phillips2023, SettySi2023b, Fabrizio2023, Goldman2023} have considered the role of Green's function zeros on topological winding numbers. For example, since the real part of Green's function changes sign across the surface of zeros, it can be shown that they contribute to the
Luttinger count~\cite{Dzyaloshinskii2003, AGD, Yunoki2017}. Along the same lines, Green's function zeros are essential to the understanding of index theorems~\cite{Volovik2003} and three dimensional topological indices~\cite{sym_green, Gurarie2011-2, Xu2015} in the presence of strong correlations.  
However, the manner in which these winding numbers are related to physical properties in the strong correlation limit has only recently picked up steam. In particular, the precise relationship between one- and three- dimensional topological windings ($N_1, N_3$ which contain contributions from Green's function zeros) and correlation functions has drawn considerable attention over the last few years. A definitive answer to this question is significant since it clarifies how Green's function zeros contribute to physical observables. \par  Unlike in the weakly interacting scenario, previous   numerical evidence in the presence of strong correlations suggested that physical correlation functions are not directly set by topological invariants~\cite{Xu2015}. Recently several other numerical and exact results have indeed definitively recognized this fact \cite{Xu2015, Phillips2023, SettySi2023b, Fabrizio2023, Goldman2023}. That this is the case can be understood by  requiring  
 that physical properties  be independent of chemical potential variations across frequencies where Green's function zeros occur within a correlated insulating gap~\cite{Rosch2007}.  
In addition, when determining the zeros' contribution to physically measurable correlations, it is crucial to 
keep track of conservation laws and the associated Ward identities. However, a naive equality between topological indices and correlation functions directly violates the above requirements. \par
Recently, some of us argued that such a violation can be resolved by the addition of backflow terms that restore conservation laws and leave physical properties invariant under chemical potential changes across the correlation gap with Green's function zeros~\cite{SettySi2023b}. Specifically, we showed that the total charge $N$ and Hall conductivity $\sigma_{xy}$ in a strongly correlated system with Green's function zeros take a form 
\beq
N &=& v_l - \delta v,  \\
\sigma_{xy} &=& N_3 + \Delta N_3. 
\eeq
Here $v_l$ is the Luttinger volume and $\delta v, \Delta N_3$ are backflow terms for the total charge and current correlator respectively. The topological invariants $v_l , N_3$ individually contain contributions from Green's function zeros; nonetheless, the backflow terms ensure that the physical properties are invariant under chemical potential variations up to the correlated gap while preserving Ward identities.  
Hence we can conclude that Green's function zeros indeed contribute to physical observables while also satisfying the above physical constraints 
that are imposed 
on general grounds~\cite{SettySi2023b}.

 \end{document}